\documentclass[12pt]{article}

\usepackage{arxiv}

\usepackage[utf8]{inputenc} 
\usepackage[T1]{fontenc}    
\usepackage{hyperref}       
\usepackage{url}            
\usepackage{booktabs}       
\usepackage{amsfonts}       
\usepackage{nicefrac}       
\usepackage{microtype}      
\usepackage{lipsum}		
\usepackage{graphicx}
\usepackage{doi}
\usepackage{scalerel,stackengine}
\usepackage{graphicx,multicol}
\usepackage{multirow} 
\usepackage{indentfirst,csquotes}
 \usepackage{array} 
\usepackage{abstract}
 \usepackage{algorithm2e}
\usepackage[caption=false]{subfig}
\usepackage{amssymb,amsthm,amsmath}
\usepackage{xcolor,paralist,hyperref,fancyhdr,etoolbox}
\newtheorem{theorem}{Theorem}
\newtheorem{definition}[theorem]{Definition}

\usepackage[round]{natbib} 

\usepackage[utf8]{inputenc}
\numberwithin{equation}{section}

\newtheorem{rmk}{Remark}
\usepackage{hyperref}

\hypersetup{ colorlinks=true, linkcolor=black, filecolor=black, urlcolor=black,citecolor = blue }

\usepackage{lipsum}
\usepackage{multicol}

\usepackage{listings}
\title{A geometric approach in non-parametric Changepoint detection in circular data}

\author{ { Surojit Biswas}\thanks{ webpage: https://sites.google.com/view/surojitbiswas/home?authuser=0} \\
	Department of Mathematics\\  IIT Kharagpur, India-$721302$ \\
	\texttt{surojit23@iitkgp.ac.in} \\
	\And
    {Buddhananda Banerjee}\thanks{ webpage: https://sites.google.com/site/buddhanandastat/} \\
	Department of Mathematics\\  IIT Kharagpur, India-$721302$ \\
	\texttt{bbanerjee@maths.iitkgp.ac.in } \\
	 \AND
     {Arnab Kumar Laha} \\
	Operations and Decision Science Department \\  IIM Ahmedabad,
    India-380015 \\
	\texttt{arnab@iima.ac.in } \\
}

\date{}

\renewcommand{\shorttitle}{Changepoint ditection in Cyclone Data}

\hypersetup{
pdftitle={A template for the arxiv style},
pdfsubject={q-bio.NC, q-bio.QM},
pdfauthor={David S.~Hippocampus, Elias D.~Striatum},
pdfkeywords={First keyword, Second keyword, More},
}

\begin{document}
\maketitle
\begin{abstract}

In many temporally ordered data sets, it is observed that the parameters of the underlying distribution change abruptly at unknown times. The detection of such changepoints is important for many applications.  While this problem has been studied substantially in the linear data setup, not much work has been done for angular data.  In this article, we utilize the intrinsic geometry of a torus to propose new non-parametric tests. First, we propose new tests for the existence of changepoint(s) in the concentration, and second, a test to detect mean direction and/or concentration.  The limiting distributions of the test statistics are derived, and their powers are obtained using extensive simulation. It is seen that the tests have better power than the corresponding existing tests. The proposed methods have been implemented on three real-life data sets, revealing interesting insights.  In particular, our method, when used to detect simultaneous changes in mean direction and concentration for hourly wind direction measurements of the cyclonic storm ``Amphan,” identified changepoints that could be associated with important meteorological events.
\end{abstract}


\section{Introduction}
Angular data ( a.k.a. directional data or circular data)  arises from the study of phenomena that are circular or periodic. In contrast to linear data, there is no universally accepted ordering of circular data for its entire range. The choice of the zero direction and the clockwise or anti-clockwise orientation play a critical role while working with circular data. Examples of circular data include angular measurements such as active galactic nuclei (astronomy), dihedral angles of protein structure (bio-informatics), wind direction (meteorology), sea wave direction (climatology), etc. The reader can refer to \cite{Mardia_2000} for a more in-depth exploration of circular data.\
 
Changepoint analysis is an important statistical concept to detect unexpected shifts or alterations within a temporally ordered data set.   These changes might be caused by a change in the parameter(s) within the same family of distributions or a complete change in the family of distributions. The usual statistical analysis of data gets heavily affected because of the existence of changepoints in the data set. Therefore, the primary goal of changepoint analysis is to conduct a statistical test to check for the existence of a changepoint in a given sequence of data. 
A substantial amount of research in changepoint analysis has been carried out for real-valued random variables   \cite[see][]{fearnhead2019changepoint,haynes2017computationally,Hovarth_1999,Antoch_1997,Cobb_1978, Davis_1995}, vector-valued random variable \cite[see][]{Kirch_2014,Kokoszka_2000, shao2010testing, anastasiou2023generalized,pishchagina2023online}, and functional valued random variable \cite[see][]{horvath2012inference, banerjee2018more, 
Horman_2010, banerjee2020data}. 

In the context of angular data, there has been limited exploration of the changepoint problem. The changepoint in angular data may occur in the mean direction, concentration, or both. Mathematically,
Let $\theta_1, \theta_2, \ldots, \theta_n \in [0,2\pi)$ be angular random variables. A changepoint is defined as a fixed and unknown point $k^* \in \{ 0,1, \ldots,n\}$ such that $\theta_1, \theta_2, \ldots, \theta_{k^*}$ follow a distribution $F(\theta;\lambda_1)$, and $\theta_{k^*+1}, \theta_{k^*+2}, \ldots, \theta_n$ follow a different distribution $F(\theta;\lambda_2)$, where $\lambda_1 \neq \lambda_2$, and $F(\theta;\lambda_1) \neq F(\theta;\lambda_2) \mbox{~~for all~~} \theta \in [0,2\pi)$.
For the first time, \cite{lombard1986change} introduced a pioneering rank-based test to detect changepoints in the location change and change in concentration parameter for angular data. Following this work, \cite{grabovsky2001change} put forth a modified CUSUM procedure for testing the change in concentration parameter of the angular distribution.
\cite{ghosh1999change},  proposed a likelihood-based approach for addressing changepoint detection in the mean direction for the von Mises distribution only. Additionally, \cite{sengupta2008likelihood} introduced a novel likelihood-based method, referred to as the likelihood integrated method.  These collective efforts mark significant strides in developing methodologies to detect changepoints within the domain of angular data. 

In this article, we deal with the changepoint problems of angular data in concentration and/or mean direction and separately in concentration. 
Applying the notion of the ``square of an angle"  introduced by \citet{biswas2025semi}, we propose a test for identifying the changepoint(s) in the concentration of angular data. We also obtain the asymptotic pivotal distribution of the test statistic under the null hypothesis of no change, which is free from the concentration as well as the mean direction (which is a nuisance parameter). We will refer to this test as the \textit{Square Angle Concentration Change  (SACC)} test in the rest of the paper.
An extensive simulation study shows that the SACC test is more powerful compared to the existing one by \cite{grabovsky2001change}.
Furthermore, using the notion of  `square of an angle,' we propose a new test for detecting changepoint(s)  for the mean direction and/or the concentration in general.  We will refer to this test as the \textit{Square Angle General Change  (SAGC)} test in the rest of the paper. 
Here as well, we obtain the asymptotic pivotal distribution of the test statistic under the null hypothesis of no change.

We illustrate the method for detecting changepoints in concentration on the Acrophase data set \cite[see][]{lombard2017sequential} related to systolic blood pressure (SBP) in a patient undergoing clinical depression episodes.   The results reveal that the proposed methods effectively identify changepoints in concentration. Lastly, we have executed the method for detecting changepoints in the concentration and/or mean direction on a weather data set \cite[see][]{data2023} of a super cyclonic storm ``Amphan”  over the Bay of Bengal that severely hit and caused extensive damages in eastern India and Bangladesh during the period 16th May 2020 to 21st May 2020. A newly developed exploratory graphical data analysis tool \textit{Circular temporal plot} (see Section \ref{ct_plot} for a description of the plot) is used to guide the presence of changepoints in a temporally ordered data set of angular observations.

The flow of the work is given below.
We breafly study the intrinsic geometry of the curved torus in Section \ref{int_geo_torus}. Section \ref{changepoint detection} is dedicated to changepoint detection in concentration with known mean direction (Section \ref{section_cp_comcentration}), and changepoint detection in either of the parameters (Section \ref{section_cp_general}) when both are unknown. 
Section \ref{simulation} presents a comparison of the proposed method with existing methods for different circular distributions, namely, von Mises, Kato-Jones, and Wrapped Cauchy. In Section \ref{data_analysis}, we implement the proposed methods for changepoint detection in concentration for Acrophase data (Section \ref{acrophase}), and in both for Amphan super cyclone data (Section \ref{Amphan}). 
The concluding Section \ref{conclusion} is followed by the algorithms and  some added diagrams and tables in Supplementary material.

\section{Intrinsic geometry of torus}
 \label{int_geo_torus}
 Before getting into the proposed methods, we briefly discuss some basic tools from Riemannian geometry.  Here, we introduce a few definitions of tangent space, the first fundamental form, and the area element. The reader may see  \cite{Gallier_2020} for details.

 \begin{definition} Let $\mathcal{M}$ be a Riemannian surface. Then the set of all tangent vectors $v$ at $x\in \mathcal{M}$ is called the \textbf{tangent space} to the point $x$ and it is denoted by $T_{x}\mathcal{M}.$
 \label{tangent space}
 \end{definition}
Let  $\mathcal{M} \subset \mathbb{R}^3$ be a  Riemannian surface defined by $X:\mathbb{R}^2 \rightarrow \mathcal{M}.$ Then a curve $\gamma(t)$ on $\mathcal{M}$ parametrized by $t\in [a,b]$ can be defined as $\gamma(t)=X(u(t),v(t)).$  Therefore, the velocity vector can be obtained as
$$\gamma'(t)=\dfrac{\partial X}{\partial u}\frac{du}{dt}+\frac{\partial X}{\partial v}\frac{dv}{dt}=[X_u, X_v] [u',v']^T$$ 
Thus, we can represent the velocity vector as the linear combination of the basis vectors $X_u=\dfrac{\partial X}{\partial u}$ and $X_v=\dfrac{\partial X}{\partial v}$, with coefficients $u'=\dfrac{du}{dt}$ and $v'=\dfrac{dv}{dt}.$
Let $s(t)$ be the arc length along $\gamma$ with $s(a)=0$ then 
$s(t)= \int_{a}^{t} ||\gamma'(r)|| \,dr ,$ so, we have 
$\dfrac{ds}{dt}=||\gamma'(t)|| .$ Now,  the \textbf{ first fundamental form} or \textbf{metric form} of the surface $ \mathcal{M}$ can be obtained as
\begin{eqnarray}
    \left( \dfrac{ds}{dt}  \right)^2&=&  \langle  \gamma'(t),\gamma'(t) \rangle \nonumber\\
    &=&\langle (u'X_u+v'X_v),(u'X_u+v'X_v) \rangle \nonumber\\ 
     &=&(u')^{2}\langle X_u , X_u\rangle +2u'v'\langle X_u,X_v\rangle +(v')^2\langle X_v , X_v\rangle \nonumber\\
     &\equiv& (u')^{2}~E +2u'v'~F+(v')^2~G \nonumber\\
     &=& \begin{bmatrix}
           u' \\
           v'  
          \end{bmatrix}^T 
          \begin{bmatrix}
           E & F \\
           F & G 
          \end{bmatrix} 
          \begin{bmatrix}
           u'  \\
           v' 
          \end{bmatrix}\nonumber
\end{eqnarray}
where $E=\langle X_u , X_u\rangle,$ $F=\langle X_u, X_v\rangle$, and $G=\langle X_v , X_v\rangle,$ with usual  inner-product  $\langle \cdot,\cdot\rangle$.  
\begin{definition}
  The area element, $dA$ of the surface $\mathcal{M}$ determined by  $X(u,v)$ is defined by 
  $$dA=|X_u \times X_v|~ dudv= \sqrt{EG-F^2} ~dudv.$$
  Hence, the total surface area of the surface   $\mathcal{M}$  is

  $$ A=\int \int dA~dudv=\int \int \sqrt{EG-F^2} ~dudv.$$
  \label{general area element}
\end{definition}

 
The rest of our work will be based on the curved torus defined by the parametric equation 
\begin{equation}
  X(\phi,\theta)=\{  (R+r\cos{\theta})\cos{\phi}, (R+r\cos{\theta})\sin{\phi}, r\sin{\theta} \}\subset \mathbb{R}^3, 
  \label{torus para equn}
\end{equation}
with the parameter space $\{ 
 (\phi,\theta):0<\phi,\theta<2\pi\}= \mathbb{S}_1 \times \mathbb{S}_1,$ known as  flat torus. 
 Now, the partial derivatives of $X$ with respect to $\phi$, and $\theta$ are 
$$ X_{\phi}=\{ -(R+r\cos{\theta})\sin{\phi}, (R+r\cos{\theta})\cos{\phi}, 0 \}$$ and
$$X_{\theta}=\{  -r\sin{\theta} \cos{\phi}, -r\sin{\theta}\sin{\phi}, r\cos{\theta}  \}$$ respectively. Hence, the coefficients of the first fundamental form are
 \begin{equation}
 \begin{aligned}
     E=\langle X_{\phi},X_{\phi}\rangle &= (R+r\cos{\theta})^2\\
    F=\langle X_{\phi},X_{\theta}\rangle &= 0 \\
     G=\langle X_{\theta},X_{\theta}\rangle &= r^2
 \end{aligned}
 \label{fff cof}
 \end{equation}
leading to the area element of the curved torus (Equation-\ref{torus para equn}) 
\begin{equation}
    dA=r(R+r\cos{\theta})~d\phi~d\theta
    \label{torus area element}
\end{equation}
from Definition-\ref{general area element}. 

\begin{figure}[h!]
    \centering
    \subfloat[]{%
        {\includegraphics[width=0.4\textwidth, height=0.31\textwidth]{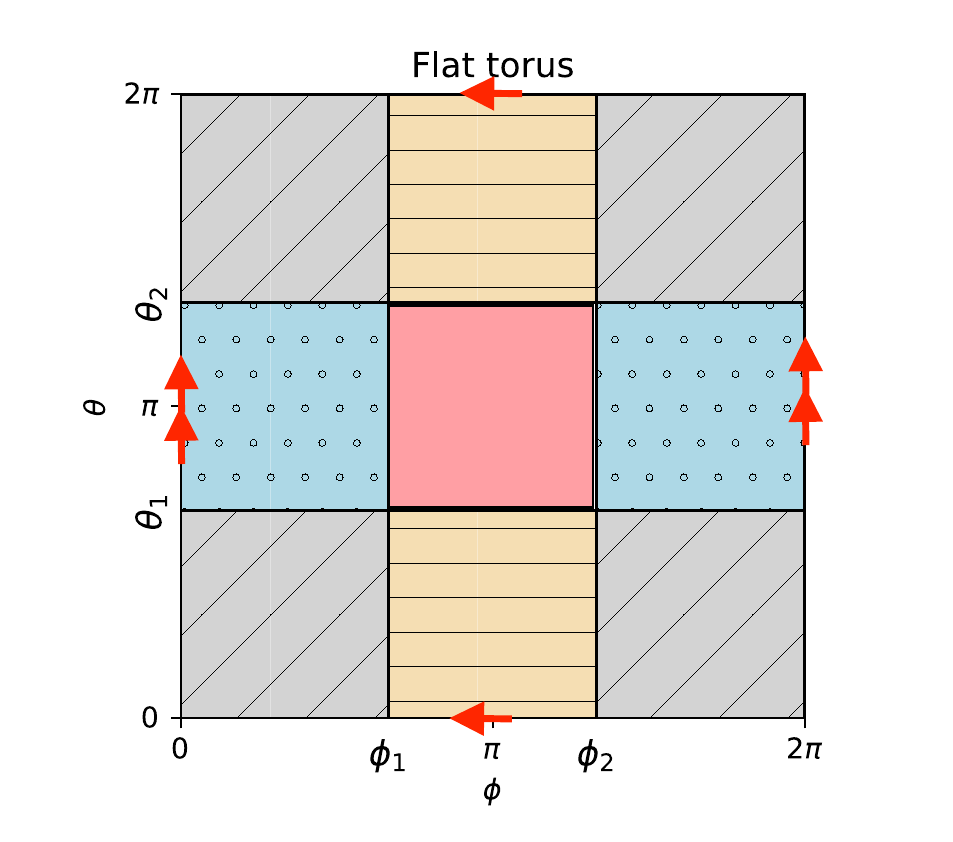}}}\hspace{5pt}
          \subfloat[]{%
        {\includegraphics[width=0.4\textwidth, height=0.31\textwidth]{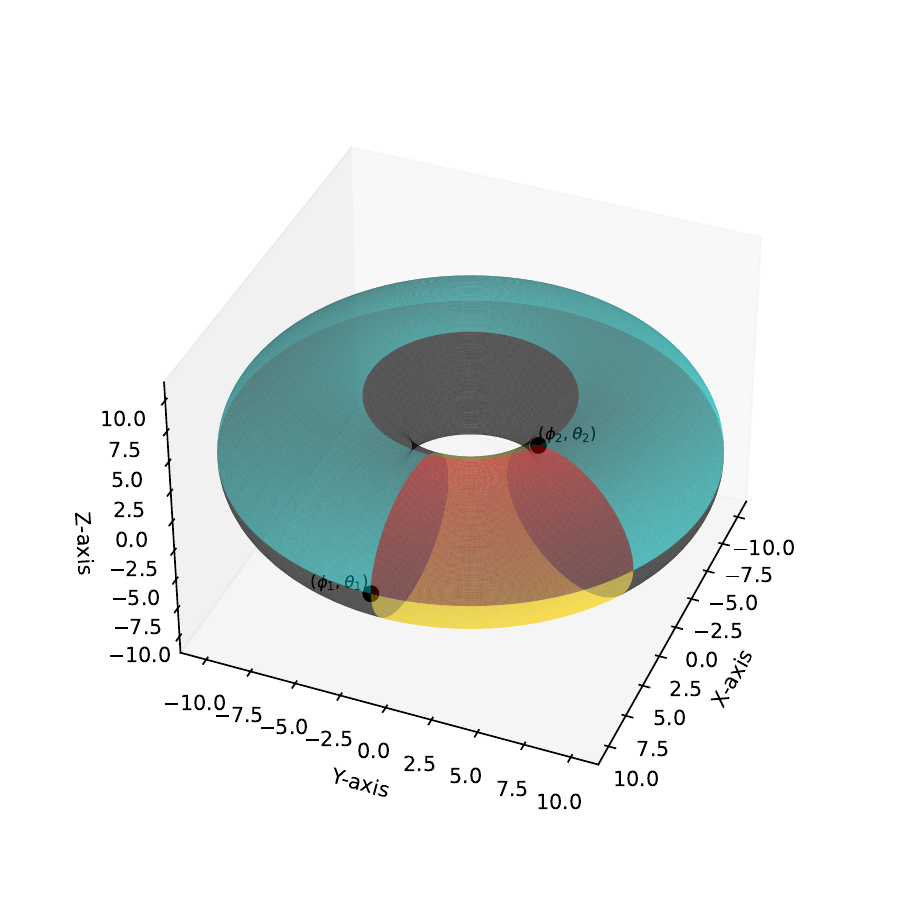}}}\hspace{5pt}

    \caption{(a) Area between $(\phi_1, \theta_1)$, and $(\phi_2, \theta_2)$ on flat torus. (b) Area between $(\phi_1, \theta_1)$, and $(\phi_2, \theta_2)$ on curved torus.}
    \label{area_plot}
\end{figure}

Let $\phi$ and $\theta$ denote the horizontal and vertical angular coordinates of a torus, respectively, with each ranging over $[0, 2\pi)$. To begin, we consider the surface area bounded by the angular coordinates $(0,0)$ and $(\phi, \theta)$ on the curved torus defined in Equation-\eqref{torus para equn}. These two points, when considered on the flat torus (i.e., the domain $[0, 2\pi) \times [0, 2\pi)$), correspond to a rectangular region in parameter space. When this region is mapped to the surface of the curved torus, the corresponding surface area can be computed using the area element in Equation-\eqref{torus area element}.
It is important to note that the curved torus surface, with respect to the pair $(0, 0)$ and $(\phi, \theta)$, is naturally partitioned into four mutually exclusive and exhaustive regions. These partitions of the flat torus domain are given by:
$$
\begin{aligned}
\mathbb{T}_1 &=& [0, \phi] \times [0, \theta], 
\mathbb{T}_2 &=&[\phi, 2\pi] \times [0, \theta], \\
\mathbb{T}_3 &=& [0, \phi] \times [\theta, 2\pi], 
\mathbb{T}_4 &=& [\phi, 2\pi] \times [\theta, 2\pi].
\end{aligned}
$$
These regions, once mapped through the surface parametrization in Equation-\ref{torus para equn}, produce distinct areas on the curved torus, as visualized in Figure-\ref{area_plot}. The surface areas corresponding to these regions are denoted $A_1, A_2, A_3,$ and $A_4$, respectively. Now, following the work of \cite{biswas2025semi}, each area is computed via the integral:
$
A_i = \iint_{\mathbb{T}_i} dA(s, t), \quad \text{for } i = 1, 2, 3, 4,
$
where the area element is derived in Equation-\eqref{torus area element}.

Analogous to the notion of circular distance - which is the length of the smaller arc between two angles, and the notion of geodesic distance on a surface - which is the length of the shortest path joining two points on the surface, \cite{biswas2025semi} have defined the \textit{proportionate area} included between these points $(0,0)$, $(\phi,\theta)$ as given   Definition-\ref{torus area of a segemnt} and 
\textit{the square of an angle} $\theta$ as given in Definition-\ref{zero centered circle area of a segemnt} below. 
\begin{definition}
	The \textit{proportionate area included between the   $(0,0)$, and $(\phi,\theta)$ } is defined as 
	
	$$A_T\left[(0,0),(\phi,\theta)\right]=\frac{\min \{A_1,A_2,A_3,A_4\}}{4\pi^2rR}.
	\label{torus area of a segemnt}
	$$
\end{definition}

\begin{definition} \textit{The square of an angle} $\theta$ is defined as
	$$A_C^{(0)}(\theta)=A_T^{(0)}(\theta,\theta)=A_T\left[(0,0),(\theta,\theta)\right], \mbox{~where~} \frac{r}{R}=1.$$
	\label{zero centered circle area of a segemnt}
\end{definition}

 The CUSUM technique \cite[see][]{grabovsky2001change} is used to construct the test statistic identifying the changepoint in the concentration.
Using the concept of `square of an angle' we now propose a test statistic to test the existence of a changepoint in the concentration,  $[{E^2(\cos \Theta)+E^2(\sin \Theta)}]^{1/2}$.
A similar idea has been extended to construct a test in concentration
and/or mean direction $\left[\tan^{-1*} \left(\frac{E(\sin \Theta)}{E(\cos \Theta)}\right)\right],$ in a sequence of angular data, 
 where $\tan^{-1*}$ is the quadrant specific inverse  is defined as  $$\tan^{-1*}(y, x) =
	\begin{cases}
		\arctan\left(\frac y x\right) &\text{if } x > 0, y \geq 0 \\
		\arctan\left(\frac y x\right) + \pi &\text{if } x < 0 , \\
		\arctan\left(\frac y x\right) + 2\pi &\text{if } x \geq 0 \text{ and } y < 0, \\
		+\frac{\pi}{2} &\text{if } x = 0 \text{ and } y > 0, \\
		\text{undefined} &\text{if } x = 0 \text{ and } y = 0.
	\end{cases}$$

\cite[see][p13, Equation-1.3.5]{jammalamadaka2001topics}.

\section{Changepoint detection}
 \label{changepoint detection}
\subsection{Changepoint detection for concentration }
\label{section_cp_comcentration}

  Let $\theta_1, \theta_2,\ldots, \theta_n \in [0,2\pi)$ be independent angular random variables.
 We consider the following testing problem : 

\begin{eqnarray}
     H_{0c} &:& \theta_{i}\overset{\mathrm{i.i.d.}}{\scalebox{1.5}{$\sim$}} F(\theta;\kappa_1) 
     \mbox{~~for all~~} i =1,2,\ldots n, \nonumber\\
     H_{1c}&:&   \begin{cases}
              \theta_{i} \overset{\mathrm{i.i.d.}}{\scalebox{1.5}{$\sim$}} F(\theta;\kappa_1) & \text{, } 1 \leq i \leq k^*\\
         \theta_{i}\overset{\mathrm{i.i.d.}}{\scalebox{1.5}{$\sim$}} F(\theta;\kappa_2)   & \text{, } (k^*+1) \leq i \leq n,
    \end{cases}
    \label{concentration test}
\end{eqnarray}
where $F$ is a circular distribution with known mean direction $\mu$, common for all $\theta_i$, and concentration  $\kappa_1 \neq \kappa_2$ under $H_{1c}$.

Let $a_i=A_C^{(0)}[(\theta_i-\mu) \mod 2\pi]\mbox{~for~} i=1,2,\ldots,n.$ Note that, $a_i$'s  are i.i.d real valued random variables under the null hypothesis, $H_{0c}$  with variance $$\sigma_{a}^2=Var(a_1)\widehat{=}\frac{1}{n-1} \sum_{i=1}^{n}\left(a_i-\Bar{a}\right)^2=\widehat\sigma_a^2,$$ where $n\bar{a}=\displaystyle \sum_{i=1}^{n}a_i.$  Now, we define a CUSUM process as
\begin{equation}
       T(k)=\frac{1}{n~\widehat\sigma_a^2 }\left[ \sum_{i=1}^{k} a_i-k\Bar{a}   \right]^2  \mbox{~~for all~~} k=1,\ldots,n 
       \label{cusum process}
\end{equation}
to construct the test statistic
\begin{equation}
    \Lambda_n=\displaystyle \max_{1 \leq k < n}   \frac{T(k)}{\sqrt{\frac{k}{n}\left(1-\frac{k}{n} \right)}}.
    \label{concentration_test_statistic}
\end{equation}

Here, we reject the null hypothesis, $H_{0c}$, if $\Lambda_n>l_{\alpha}$, where, $l_{\alpha}$ is the upper $\alpha$ point of the exact (or asymptotic) distribution of $\Lambda_n$. The closed-form distribution of $\Lambda_n$ is not available; hence, we need to take recourse to simulation to obtain the cut-off value $l_{\alpha}$. But, when $n$ is large, the limiting distribution of $\Lambda_n$ can be derived as follows.

Let us consider $u \in (0,1)$, and denote $k= \lfloor{nu}\rfloor$. Hence, from Equation-\ref{cusum process} we can write

\begin{equation}
    T_{n}(u)=T(\lfloor{nu}\rfloor)=\left( \frac{1}{\sqrt{n~} \hat\sigma_{a}} \left[ \sum_{i=1}^{\lfloor{nu}\rfloor} a_i-u \sum_{i=1}^n a_i   \right]\right)^2.
    \label{concentration asym}
\end{equation}

Then, with the proper embedding of Skorohod topology in $D[0,1]$ \cite[see][Ch. 3]{billingsley2013convergence}, under the null hypothesis, $H_{0c}$, and as $n \uparrow \infty$, the process $T_n(u)$  converges weakly to $B_{0}^2(u),$ where $B_{0}(u)$ is the standard Brownian bridge on $[0,1].$ As a consequence 
\begin{equation}
    \Lambda_n  \overset{d}{\to} \displaystyle \sup_{0<u<1} B_{0}^2(u)=B_\infty, \mbox{~~say}.
    \label{bbridge}
\end{equation}

Hence, we can compute the upper-$\alpha$ value, $l_{\alpha},$ from the above limiting distribution of the test statistic (Equation-\ref{bbridge}). When the common mean direction ($\mu$)  of the data is unknown, it can be estimated by the circular mean ($\hat{\mu}$) of the entire sample, and the $a_{i}'s$ in Equation-\ref{cusum process} can be replaced by 
$$\hat{a}_i=A_C^{(0)}[(\theta_i-\hat{\mu}) \mod 2\pi]\mbox{~for~} i=1,2,\ldots,n$$
to define $\widehat{T}(k)$ which will provide a  test statistic
similar to $\Lambda_n$, namely,
\begin{equation}
    \widehat{\Lambda}_n=\displaystyle \max_{1 \leq k < n}   \frac{\widehat{T}(k)}{\sqrt{\frac{k}{n}\left(1-\frac{k}{n} \right)}}.
    \label{concentration_test_statistic_estim}
\end{equation}

From Figure-\ref{fig: density_sb_null}, it can be seen that the large sample distribution of the test statistic $\Lambda_n$ under the null hypothesis, $H_{0c}$, is close to the distribution of  $B_\infty.$ Also note that when the common mean direction is unknown, the large sample distribution (see supplementary, Figure-S\ref{fig: density_sb_null_mean_unknown} of the test statistic $\widehat{\Lambda}_n$ is well approximated by that of the  $B_\infty.$  under $H_{0c}$. Hence, the cut-off value obtained from the distribution $B_\infty$ can be used to conduct an upper-tail test. Thus, for the SAAC test we
\begin{equation}
\mbox{Reject~} H_{0c} \mbox{~if~} l_\alpha< \widehat{\Lambda}_n.
    \label{test_rule_concentration }
\end{equation}
at level $\alpha.$

\subsection{Changepoint detection- The general case}  
\label{section_cp_general}

Let $\theta_1, \theta_2,\ldots, \theta_n \in [0,2\pi)$ be independent angular random variables.
 We consider the following testing problem : 

\begin{eqnarray}
     H_{0g} &:& \theta_{i}\overset{\mathrm{i.i.d.}}{\scalebox{1.5}{$\sim$}} F(\theta;\boldsymbol{\xi}_1) 
     \mbox{~~for all~~} i =1,2,\ldots n, \nonumber\\
     H_{1g}&:&   \begin{cases}
              \theta_{i} \overset{\mathrm{i.i.d.}}{\scalebox{1.5}{$\sim$}} F(\theta;\boldsymbol{\xi}_1) & \text{, } 1 \leq i \leq k^*\\
         \theta_{i}\overset{\mathrm{i.i.d.}}{\scalebox{1.5}{$\sim$}} F(\theta;\boldsymbol{\xi}_2)   & \text{, } (k^*+1) \leq i \leq n,
    \end{cases}
    \label{general test}
\end{eqnarray}
where $\boldsymbol{\xi}_1,\boldsymbol{\xi}_2 $ are  suitable  vector valued  parameters and $\boldsymbol{\xi}_1 \neq \boldsymbol{\xi}_2.$
under the alternative hypothesis $H_{1g}$.

Using the Equation-\ref{zero centered circle area of a segemnt}, we consider the corresponding non-centered and centered areas on the surface of a curved torus
$\Tilde{a}_i=\left[2\left( \delta_{(\theta_i<\pi)}-0.5 \right) \right]A_C^{(0)}(\theta_i)$ and $\hat{a}_i=A_C^{(0)}[(\theta_i-\hat{\mu}) \mod 2\pi]$ , respectively,  where   $\hat{\mu}$ is the estimated circular mean of the entire sample and $\left[2\left( \delta_{(\theta_i<\pi)}-0.5 \right) \right]$ is an associated sign to the angle $\theta_i  \mbox{~for~} i=1,2,\ldots,n.$
Now, let $ \displaystyle d_i=\max\{\hat{a}_i,\Tilde{a}_i\} \mbox{~for~} i=1,2,\ldots,n,$ and

$$S_{d}^2{=}\frac{1}{n-1} \sum_{i=1}^{n}\left(d_i-\Bar{d}\right)^2,$$ where $n\bar{d}=\displaystyle \sum_{i=1}^{n}d_i.$  Now, we define the  CUSUM process as
\begin{equation}
       U(k)=\frac{1}{n~S_{d}^2} \left[ \sum_{i=1}^{k} d_i-k\Bar{d}   \right]^2  \mbox{~~for all~~} k=1,\ldots,n 
       \label{gen cusum process}
\end{equation}
and define  the test statistic
\begin{equation}
    \mathcal{G}_n=\displaystyle \max_{1 \leq k < n}   \frac{U(k)}{\sqrt{\frac{k}{n}\left(1-\frac{k}{n} \right)}}.
    \label{general_test_statistic}
\end{equation}

Here, we reject the null hypothesis, $H_{0g}$, if $\mathcal{G}_n>g_{\alpha}$, where, $g_{\alpha}$ is the upper $\alpha$ point of the   distribution of $\mathcal{G}_n$ under the null hypothesis. 
 It can be seen from the simulation results reported in Figure-S\ref{fig: density_general_null} (see supplementary)  that the asymptotic null distribution is close to the distribution of  $B_\infty.$ Hence we propose to obtain the cut-off value $g_{\alpha}$ from the distribution of   $B_\infty$  for large $n$. 
 Thus, for the SAGC test we
\begin{equation}
\mbox{Reject~} H_{0g} \mbox{~if~} g_{\alpha}< \mathcal{G}_n.
    \label{test_rule_general}
    \end{equation} The performance of this test is studied through extensive simulation in Section-\ref{simulation}.  

\begin{rmk}
    If there exists more than one changepoint in the given sequence of the data, we can iteratively use the proposed methods to detect these changepoints.  The binary segmentation scheme may be used in which the test is applied to each segment, beginning with the full sequence, and depending on the observed p-values, re-segmentation of the data sequence is performed to detect the possible changepoints.  
    \label{rmk-binarysegment}
\end{rmk}

\section{Numerical Studies}
\label{simulation}
A comprehensive simulation study was conducted first to detect the changepoint in the concentration, then to detect the changepoint in the mean direction, and finally for both. To do so, we have considered the von Mises distribution with the probability distribution function 
\begin{equation}
    f_{vm}(\theta)=\frac{e^{\kappa\cos(\theta-\mu)}}{2\pi I_{0}(\kappa)},
\label{von mises}
\end{equation}
where $0\leq \theta<2\pi$, $0\leq \mu<2\pi$, $\kappa>0$, and $ I_{0}(\kappa)$ is the modified Bessel function with order zero evaluated at $\kappa.$ \

\begin{figure}[h!]
    \centering
        {\includegraphics[width=0.4\textwidth, height=0.35\textwidth]{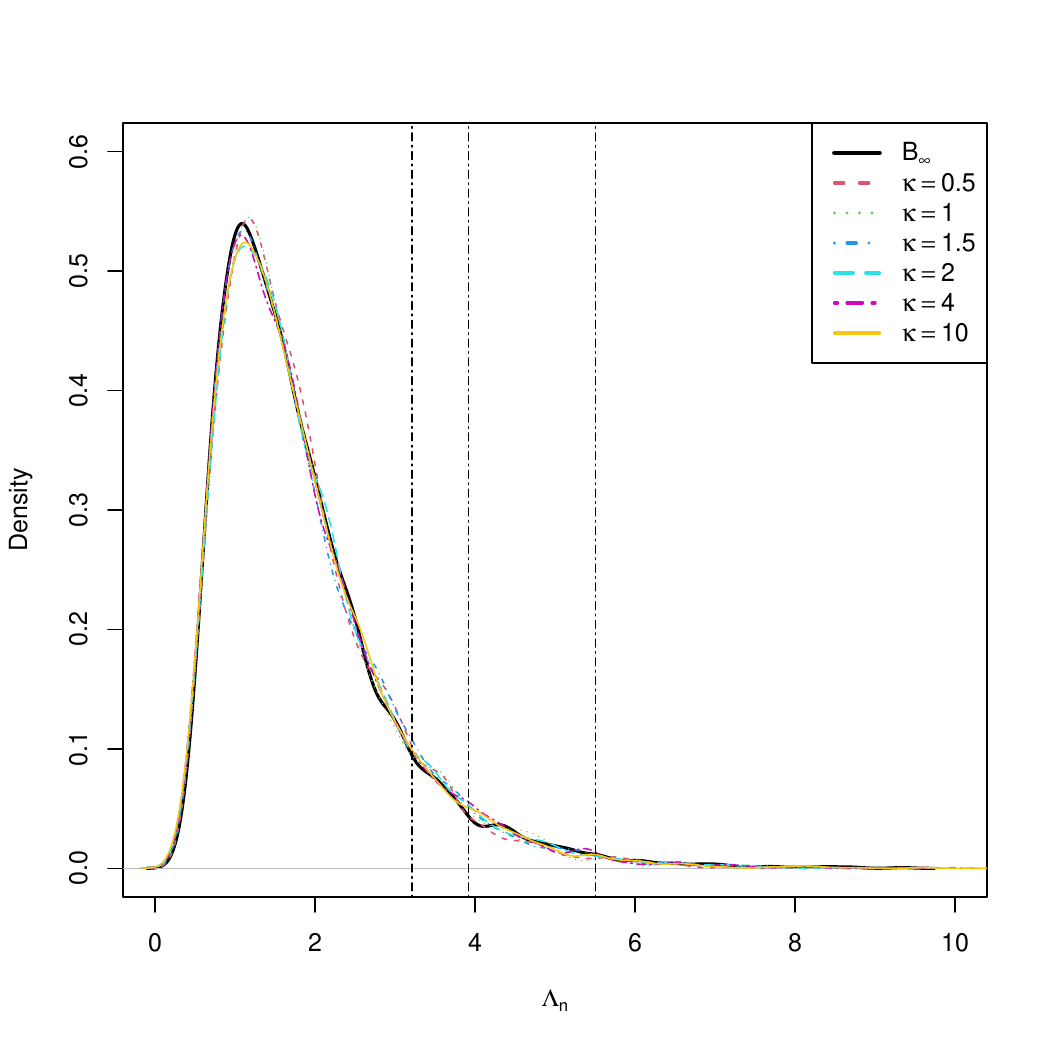}}\hspace{5pt}
  
    \caption{ Density plot of the SACC test statistic, $\Lambda_n$ under $H_{0c}$ with sample of size $n=1000$ from von Mises distribution with the mean  $\mu=0$, and different concentration parameters, $\kappa=0.5,1,1.5,2,4,10$. Density plot of the limiting random variable $B_{\infty} $ along with the $0.90th,0.95th,$ and $0.99th$ quantiles. 
 }
    \label{fig: density_sb_null}
\end{figure}

 The Figure-\ref{fig: density_sb_null} displays density plot of the  test statistic, $\Lambda_n$ under $H_{0c}$
 with the sample size of $n=1000$ from von Mises distribution with  mean  $\mu=0$, and different concentration parameters, $\kappa=0.5,1,1.5,2,4,10$. The number of iterations for each specification is conducted $5 \times 10^3$ number of times.
It is evident from Figure-\ref{fig: density_sb_null} that the densities of the test statistic for different concentration parameters are nearly identical, and these are close to the density of the limiting distribution of the random variable $B_{\infty}$ (Equation-\ref{bbridge}). 
 \cite{grabovsky2001change}  provide a test statistic  for testing $H_{0c}$ against $H_{1c}$.   They proved that the limiting distribution of the test statistic is the same as that of the random variable 
 $$\Delta_{\infty}= \displaystyle \sup_{0<t<1} \left[ \frac{B_{1}^2(t)+B_{2}^2(t)}{t(1-t)}  \right]^{\frac{1}{2}},$$ 
 where $B_{1}(t), B_{2}(t)$ are independent standard Brownian bridges on $[0,1]$ if $\mu=0$ known.

In Figure-S\ref{sb_null_kappa_plot}(A)  $\&$ S\ref{sb_null_kappa_plot}(B) (see supplementary), the distribution of the estimated location of the changepoint and the histogram of the SACC test statistic, $\Lambda_n$ (Equation-\ref{concentration_test_statistic}) are presented under the null hypothesis, $H_{0c}$, respectively.  We conducted $5 \times 10^3$ iterations using random samples from the von Mises distribution with a sample size of $n=500$, mean $\mu=0$, and concentration parameter $\kappa=1$.
In Figure-S\ref{sb_null_kappa_plot}(B), the 90th, 95th, and 99th percentiles are denoted by the vertical lines from left to right, respectively. In Table-S\ref{table:cut-off for concentration} (see supplementary) we provide the cut-off values of the test statistic $\Lambda_n$ under the null hypothesis, $H_{0c}$ when the sample of size of $50, 100,200, 500,$ and $1000$ are drawn from von Mises distribution with mean direction $\mu=0,$ and different concentrations parameters $\kappa=0.5,1,1.5 ,2,4$  and $10$. The table also provides the cut-off values from the limiting distribution of $B_\infty$ (Equation-\ref{bbridge}) with the grid size of $50, 100,200, 500,$ and $1000$, respectively, and they are denoted as $B_\infty^{(n)}$ in general.

 The power computation of the SACC test has been summarised in the Algorithm-\ref{alg:algo_power_kappa} (see supplementary). Under the alternative hypothesis, $H_{1c}$, the Figure-S\ref{sb_alt_kappa_plot}(A) $\&$ S\ref{sb_alt_kappa_plot}(B) (see supplementary)show the distribution of the estimated location of the changepoint and the histogram of the SACC test statistic, $\Lambda_n$ (Equation-\ref{concentration_test_statistic}), respectively. 
 Here also, the random samples are drawn from the von Mises distribution with a sample size of $n=500$, mean $\mu=0$. The concentration parameter is $\kappa_1=1$ before the true changepoint  $k^{*}=\frac{n}{2}$, and $\kappa_2=0.5$ after the changepoint.

 The Figures-S\ref{power_100ss}, \& S\ref{power_500ss}(see supplementary) provide a comparison of the power of the SACC test (\ref{test_rule_concentration }) with that provided by \cite{grabovsky2001change} for sample sizes of $100,$ and $500$, respectively. We consider two levels of significance, namely  $1\%,$ and $5\%$, and under the null hypothesis, $\kappa_1=2.5$ and under the alternative hypothesis, the location of changepoint is considered at $k^*=\frac{n}{2}.$ The figures are based on  $5 \times 10^3$ iterations. We see from the figures that the power of the SACC test is greater than that of the test given by \cite{grabovsky2001change}. 

The Kato-Jones distribution is a four-parameter family of circular distributions introduced in \cite{kato2010family}.  The von Mises and Wrapped Cauchy distributions are members of this general family, and some real-life applications have been found.  
The pdf of the Kato-Jones distribution is 
\begin{eqnarray}
       f_{kj}(\theta)&=&\frac{1-\rho^2}{2\pi I_{0}(k)} \exp\left[ \dfrac{\kappa \{ \xi \cos{(\theta-\eta)}-2\rho \cos \nu \}}{1+\rho^2-2\rho \cos{(\theta-\gamma)}}   \right] \nonumber\\
       && \hspace{2cm}\times \frac{1}{1+\rho^2-2\rho \cos{(\theta-\gamma)}},
    \label{kato jons}
\end{eqnarray}

where $0\leq \mu,\nu<2\pi$, and $0\leq \rho <1$, $\kappa>0$, and 
$\gamma=\mu+\nu$, $\xi=\sqrt{\rho^4+2\rho^2\cos{(2\nu)}+1}$, $\eta=\mu+\arg(\rho^2\cos{(2\nu)}+1+i\rho^2\sin{(2\nu)})$. 
Hare, $\mu$ and $ \nu$ are location $\kappa$ and $ \rho$ are concentration parameters. 

In Figures-S\ref{kjpower_kappa} \& S\ref{kjpower_r} (see supplementary) we examine the power of the SACC test (\ref{test_rule_concentration }) with that of \cite{grabovsky2001change} for the Kato-Jones distribution. We consider a sample size of $500$ and use the limiting distribution $B_\infty$ to obtain the cutoff value for the SACC test. Under the null hypothesis, $H_{0c}$, we take $\mu=\nu=0$, $\rho=0.4$, and $\kappa=2.5$.  In Figure-S\ref{kjpower_kappa} we keep $\rho$ fixed and vary $\kappa$ whereas in Figure-S\ref{kjpower_r} we keep  $\kappa$ fixed and vary $\rho$ to obtain the power functions. We observe from these figures that the SACC test (\ref{test_rule_concentration }) has substantially better power than that of \cite{grabovsky2001change}. 

 The Figure-S\ref{fig: density_general_null} (see supplementary) displays density plot of the  test statistic $ \mathcal{G}_n$ under $H_{0g}$
 the sample size of $n=1000$ the von Mises distribution with the following pairs of parameter specifications: $(\mu,\kappa)=(0,0.5),(0,1),(\frac{\pi}{6},1.5), (\frac{\pi}{6},2),(\frac{\pi}{3},4),$ $(\frac{\pi}{3},10).$
 Here we have performed $5 \times 10^3$  a number of iterations for each of the parameter specifications.
It is evident from Figure-S\ref{fig: density_general_null} that the densities for different parameter specifications are not only nearly identical but are close to the density of the limiting distribution of the random variable $B_{\infty}$ (Equation-\ref{bbridge}). 
A binary segmentation scheme (Remark \ref{rmk-binarysegment} ) for changepoint detection in either or both the parameters of von Mises distribution with a sample size of $500$ has been illustrated in Figure-S\ref{fig: general_alt_plot} (see supplementary). For a pre-specified set of existing changepoints  at $125~(\mu=\pi \mbox{~to~} \mu=\pi/2) $, $250~(\kappa=0.5 \mbox{~to~} \kappa=2) $, and $375~ (\mu=\pi/2,  \kappa=2 \mbox{~to~} \mu=\pi , \kappa=4) $ are  estimated using the proposed method of SAGC test (Algorithm-\ref{alg:algo_general}) (see supplementary)  at $129$, $238$, and $375$ respectively.

\section{Data analysis}
\label{data_analysis}
In this section, we apply the newly developed procedures to real-life data sets coming from biology, engineering, and meteorology. In Section \ref{ct_plot}, we introduce a new plotting technique for exploratory analysis of temporally ordered angular data that is helpful for changepoint analysis. In the following three Sections \ref{acrophase}, and \ref{Amphan}, we discuss the changepoint analysis of three real-life data sets and, when available, compare the findings with that of the existing procedure.    

\subsection{Circular Temporal Plot} 
\label{ct_plot}
A new scatter plot that we call the \textit{Circular Temporal Plot} is developed to visualize the temporarily ordered angular data. Suppose we have $n$ temporally ordered angular observations $\theta_1, \ldots, \theta_n$. Let us begin by considering $n$ concentric circles with the innermost circle representing time index 1, the next circle representing time index 2, etc., with the outermost circle representing time index $n$. The radius of the $i$-th circle is taken to be $\frac{i}{n}R$ where $R$ is the radius of the outermost circle. The observation $\theta_i$ is plotted on the $i$-th circle as $\frac{i}{n}R(\cos\theta_i,\sin\theta_i).$ In the presence of changepoint(s) in the mean direction, we expect to see abrupt change(s) in the direction of the plotted observations. Abrupt change(s) in the spread of the plotted observations without a change in direction is indicative of the presence of changepoint(s) in concentration. If both kinds of changes are present, we expect to see a combination of the behaviors of the above two situations in the plotted data. The use of this plot is seen to be very helpful in the preliminary analysis of the data and in identifying the appropriate technique to be used for formal analysis. The plot can be further embellished with highlighted circles that segregate the homogeneous segments within annular regions post the formal analysis of the data using an appropriate procedure. The mean direction of each homogeneous segment is plotted as a point on the highlighted circle corresponding to the outer end of the segment. A proportionate color intensity scale is used to depict the concentration of the segment. We have implemented the Circular Temporal Plot for the three data sets mentioned above (see Figure-\ref{fig: data_analysis_acro_polar}, \& \ref{fig: data_analysis_amphan_polar}).

\subsection{Acrophase Data:}
\label{acrophase}
We first analyze the Acrophase data set referred by \cite{lombard2017sequential} by our proposed method (SACC test) as well as by the method of \cite{grabovsky2001change}. In the context of circadian rhythms, the acrophase represents the time of day at which a particular rhythm reaches its peak. Systolic blood pressure (SBP) exhibits a circadian rhythm. In most people, SBP tends to be lower during periods of rest and sleep, and it increases upon waking and throughout the day. Typically, the ambulatory blood pressure monitoring (ABPM)  method is used to monitor the acrophase for SBP. Tracking the acrophase can serve as an automated early warning system for potential medical conditions before they manifest clinically.
The data set contains $306$ observations from ambulatory monitoring equipment worn by a patient experiencing episodes of clinical depression, and it is measured in radians in $[-\pi, \pi]$ that we have converted to data in $[0, 2\pi]$. 

\begin{figure*}[t]
\centering
	\includegraphics[width=0.4\textwidth,height=0.4\textwidth]{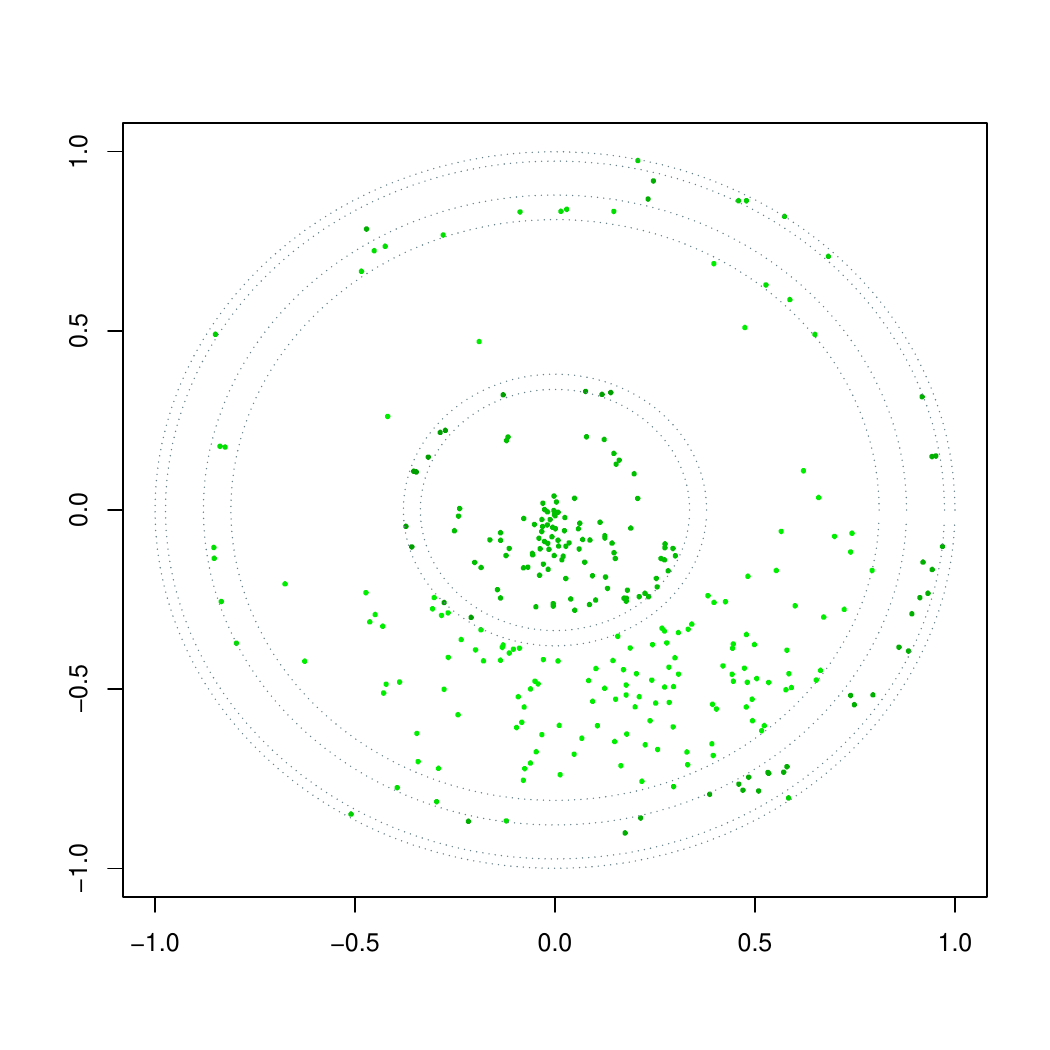}
	\caption{Scatter plot of the Acrophase data set. Six annular circles from the center to outward represent the corresponding estimated changepoints in the concentration parameter by the proposed method of the SACC test.  The color intensity scale from lower to higher has been used to depict the concentration of angular data.}
    \label{fig: data_analysis_acro_polar}
\end{figure*}

We have implemented both methods to identify the changepoint in concentration through testing. The binary segmentation procedure 
 (Remark-\ref{rmk-binarysegment}) has identified multiple changepoints. The findings and p-value have been reported in Table-\ref{table:data_concentration_table}, whereas Table-\ref{tab:concentrations value for segments} displays the estimated value of the concentration for each of the segments. The p-values have been computed with respect to the limiting distributions $B_{\infty}^{(300)}$. Note that if two successive changepoints are detected within five observations then no further analysis has been done in that short segment.
 Figure-S\ref{fig: data_analysis_plot} (see supplementary) illustrates the estimated location of the changepoints in the concentration by both methods. We also represent the data using a circular temporal plot in Figure-\ref{fig: data_analysis_acro_polar} where five annular circles from the center to outward represent the corresponding estimated changepoint in the concentration by the proposed method (SACC test).
 
\subsection{Amphan Cyclone Data:}
\label{Amphan}

The Super Cyclonic Storm (SuCS) “AMPHAN” caused extensive damage to eastern India and Bangladesh during the period 16th May 2020 to 21st May 2020  over the Bay of Bengal (BoB). As per the description by the Regional specialized meteorological center - tropical cyclones, New Delhi India Meteorological Department (see, \url{https://internal.imd.gov.in/press_release/20200614_pr_840.pdf}),  AMPHAN started forming from the remnant of a low-pressure area that occurred over the south Andaman Sea and adjoining southeast Bay of Bengal during the period  6th-12th May 2020.  This gradually developed into a well-marked low-pressure area that was observed over southeast BoB at 0300 UTC on 14th May 2020. It intensified into a cyclonic storm over southeast BoB on the  16th of May 2020. This further intensified into a SuCS over the next two and half days and maintained the intensity of SuCS for nearly 24 hours before weakening into an Extremely Severe Cyclonic Storm over west-central BoB on 19th May 2020 and made landfall at $21.65 ^\circ N, 88.3^\circ E$ on the coast of West Bengal, India during 1000-1200  UTC on 20th May, with a maximum sustained wind speed of 155 – 165 kmph gusting to 185 kmph.  
  
Intending to study the possible association of the wind direction at a chosen location with the meteorological events described above, we collected the 10-meter-above-the-sea-level wind direction data \cite[see][]{data2023} at the location with coordinates $11^{\circ} N$, and $86.5^{\circ}E$, which is about 1200 km from the location of the landfall.  The hourly data spans from May 10, 2020, 0000 UTC to May 20, 2020, 1200 UTC. This resulted in a total of 258 observations reported in degrees. 

  As discussed above, several significant meteorological events happened during the period 10th - 20th May 2020, which indicates the possibility of changepoint(s) being present in the data. Since both the mean direction and the concentration are unknown,  we executed the SAGC test described in Algorithm- \ref{alg:algo_general} ( see supplementary) to determine the existence of changepoint(s) for concentration and/or mean direction in this data set. Using the binary segmentation procedure (\ref{rmk-binarysegment}), we found the presence of multiple changepoints. Table-\ref{table:data_amphan_cp_table} reports the results, whereas Table-\ref{table:data_amphan_mean_concentration_table} shows the estimated values of the concentration and mean direction for each segment. It may be noted that we used the limiting distributions $B_{\infty}^{(258)}$,  to obtain the corresponding p-values. Also, note that if two successive changepoints are detected in a short range  (less than 24 hours) then further analysis has not been carried out in that short segment.

\begin{figure*}[h!]
\centering
	\includegraphics[width=0.4\textwidth,height=0.4\textwidth]{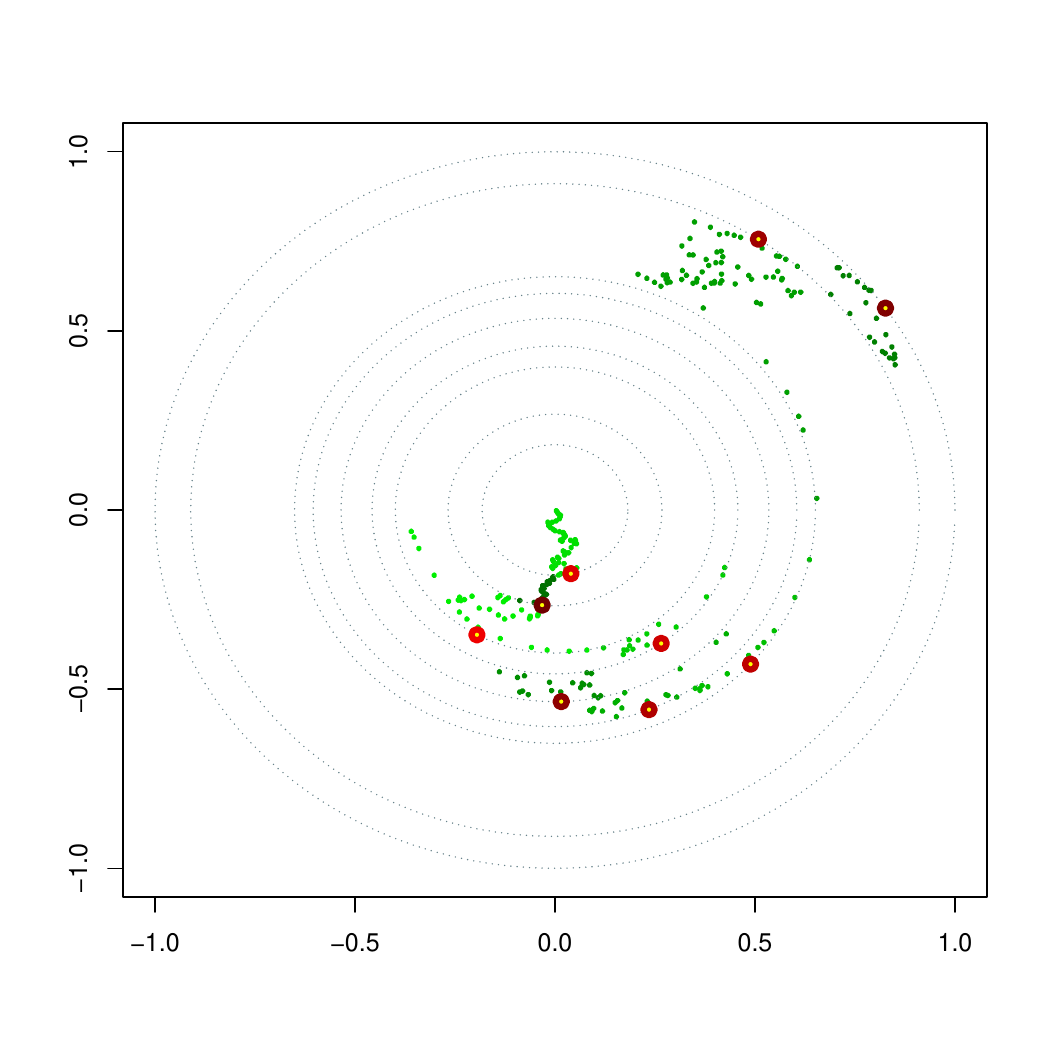}
	\caption{ Scatter plot of the data set of the Super Cyclonic Storm (SuCS) ``AMPHAN''. Eight annular circles from the center to outward represent the corresponding estimated changepoints in the mean direction and/or concentration by the proposed method of the SAGC test. The segment-wise estimated mean is represented by red bubble plot at the outer end of the corresponding segment.  The color intensity scale from lower to higher is proportionate to the concentration of angular data.}
    \label{fig: data_analysis_amphan_polar}
\end{figure*}

It is interesting to observe that the changepoints detected at locations 47 (1730 UTC, 11 May 2020) and 69 (1530 UTC, 12 May 2020) correspond to the manifestation of a low-pressure area over the South Andaman Sea and adjoining southeast BoB. The changepoint at location 103 (0130 UTC, 14 May 2020) is the identification of a well-marked low-pressure area over southeast BoB reported at 0300 UTC on 14th May 2020. Moreover, changepoints at locations 156 (0630 UTC, 16 May 2020) and 168 (1830 UTC, 16 May 2020) correspond to the occurrence of a deep depression reported at 0900 UTC on 16th May, which subsequently intensified into a cyclonic storm later in the same day.  The final changepoint was detected before landfall at location 235 (1330 UTC, 19 May 2020). Thus, we see that the estimated locations of these changepoints closely correspond to meteorological phenomena associated with the Super Cyclonic Storm. Figure-S\ref{fig: data_analysis_amphan_plot}  (see supplementary)depicts the changepoint locations estimated by the proposed method  (Algorithm- \ref{alg:algo_general}). We also represent the data using a circular temporal plot in Figure-\ref{fig: data_analysis_amphan_polar} where eight annular circles from the center to outward represent the corresponding estimated changepoints in the mean direction or/and concentration by the proposed method. The segment-wise estimated mean is represented by red bubble plot at the outer end corresponding segment.

\section{Conclusion}
\label{conclusion}
This study examines the significant yet somewhat unexplored issue of changepoint identification in angular data, focusing on changes in concentration and mean direction.  In contrast to real-valued or vector-valued contexts, angular data pose distinct issues owing to their intrinsic circular characteristics, making conventional changepoint detection techniques ineffective or inadequate.

 To address these issues, we implemented two innovative testing processes based on the newly established idea of the “square of an angle” \citep{biswas2025semi}.  The Square Angle Concentration Change (SACC) test is designed to identify changepoints in the concentration parameter of angular data.  This test produces an asymptotically key test statistic with a distribution independent of both the mean direction (a nuisance parameter) and the underlying concentration, making it particularly advantageous in real scenarios where these parameters are unknown or challenging to estimate.  The simulation findings clearly illustrate the superiority of the SACC test regarding power in comparison to the established CUSUM-based technique by \citet{grabovsky2001change}.

In addition, we proposed the Square Angle General Change (SAGC) test to handle changepoint detection in both the mean direction and the concentration simultaneously. This test, like the SACC, enjoys an asymptotically pivotal distribution under the null hypothesis, offering a robust semi-parametric framework for broader changepoint analysis in circular settings.

We validated the theoretical features of the proposed tests and illustrated their practical utility through comprehensive simulation studies.  The implementation of our methodology on two real-world datasets, the Acrophase dataset and the Amphan cyclone wind direction data demonstrates the efficacy of our approach.  In both instances, the tests effectively identified significant changepoints, providing novel insights into biological and meteorological events, respectively.  We also created an innovative exploration tool, the Circular Temporal Plot, which facilitates visual detection and corroborates the formal test results.

\section{Competing interests}
No competing interest is declared.



\section{Acknowledgement} The author, S. Biswas, acknowledges and appreciates the financial assistance provided in the form of a junior/senior research fellowship by the Ministry of Human Resource and Development (MHRD) and IIT Kharagpur, India. Author B. Banerjee would like to thank the Science and Engineering Research Board (SERB), Department of Science \& Technology, Government of India, for the MATRICS grant (File number MTR/2021/000397)  for the funding of the project.

\bibliographystyle{unsrtnat}
\bibliography{buddha_bib}  

\begin{table}[b]
\begin{center}
{\renewcommand{\arraystretch}{.5}
\begin{tabular}{|>{\centering\arraybackslash}p{1.8cm}|>{\centering\arraybackslash}p{1.8cm}|>{\centering\arraybackslash}p{1.8cm}|p{0.5cm} |>{\centering\arraybackslash}p{1.8cm}|>{\centering\arraybackslash}p{1.8cm}|>{\centering\arraybackslash}p{1.8cm}|}
    \hline
    \multicolumn{3}{|c|}{\textbf{SACC Method}} & \multicolumn{1}{c|}{} & \multicolumn{3}{|c|}{\textbf{by  \cite{grabovsky2001change}}} \\
    \cline{1-3} \cline{5-7}
   Data segment & Estimated changepoint & P-value & & Data segment & Estimated changepoint & P-value \\
   \cline{1-3} \cline{1-3} \cline{1-3} \cline{5-7}  \cline{5-7}  \cline{5-7} \cline{5-7}                               
   
 1-306 & 248 & 0.0000 &       &1-306 & 242 & 0.000\\
 1-248 & 116 & 0.0000 &       &1-242 & 126 & 0.0004 \\
1-116 & 103 & 0.0000&         &1-126 & 103 & 0.0009 \\
 1-103 & 76 & 0.1762 &        &1-103 & 87 & 0.0020 \\
104-116 & 105 & 0.0000 &      &1-87 & 59 & 0.1902\\
117-248 & 149 & 0.9593&       &88-103 & 91 & 0.9606 \\
249-306 & 269 & 0.0000 &      &104-126 & 110 & 0.0036\\
249-269 & 264 & 0.4814 &     &111-126& 115 & 0.0290\\
270-306 & 298 & 0.0372 &     &116-126  & 124 & 0.9561\\
270-298 & 281 & 0.5496 &     &127-242 & 171 & 0.6903 \\
299-306 & 302 & 0.9457 &     &243-306 & 269 & 0.0126\\
        &     &        &     &270-306 & 298 & 0.0518 \\
    \hline
\end{tabular}
}
\end{center} 
\vspace{0.3cm}
\caption{Results from the binary segmentation scheme to detect changepoint in concentration of the Acrophase data.}
\label{table:data_concentration_table}
\end{table}
\begin{table}[b]
\centering
{\renewcommand{\arraystretch}{.1}
\begin{tabular}{| >{\centering\arraybackslash}m{2.5cm} || 
                >{\centering\arraybackslash}m{1.4cm} | 
                >{\centering\arraybackslash}m{1.4cm} | 
                >{\centering\arraybackslash}m{1.4cm} | 
                >{\centering\arraybackslash}m{1.4cm} | 
                >{\centering\arraybackslash}m{1.4cm} | 
                >{\centering\arraybackslash}m{1.4cm} |}
 \hline
 Homogeneous segment & 1-103 & 104-116  & 117-248 & 249-269 & 270-298 & 299-306\\
 \hline \hline
 Estimated concentrations  & 0.5598  & 0.6288 &  0.7602 & 0.3799 & 0.7298 & 0.4391 \\
\hline
\end{tabular}
}
\vspace{0.3cm}
\caption{ Values of the concentration for each segment where no changepoint is detected by the SACC tes for the  Acrophase data.}
\label{tab:concentrations value for segments}
\end{table}

\begin{table}[t]
\centering
\renewcommand{\arraystretch}{0.5} 
\begin{tabular}{ |>{\centering\arraybackslash}p{2.5cm}|>{\centering\arraybackslash}p{2.5cm}|>{\centering\arraybackslash}p{2.5cm}| }
  \hline
Data  segment & Estimated changepoint & P-value \\
  \hline
  1-258 & 235 & 0.0000 \\
  1-235 & 103 & 0.0000\\
  1-103 & 47 & 0.0000 \\
  1-47 & 7 & 0.0008 \\
  8-47 & 26 & 0.0626 \\
  48-103 & 69 & 0.0034 \\
  70-103 & 99 & 0.1158 \\
  104-235 & 168 & 0.0000 \\
  104-168 & 156 & 0.0006 \\
  104-156 & 118 & 0.0006 \\
  119-156 & 138 & 0.0002 \\
  169-235 & 173 & 0.0648 \\
  \hline
\end{tabular}
\vspace{0.3cm}
\caption{ Results from the binary segmentation scheme to detect changepoints by the SAGC test for the  Amphan cyclone data.}
\label{table:data_amphan_cp_table}
\end{table}
\begin{table}[b]
\centering
\renewcommand{\arraystretch}{.7} 
\begin{tabular}{ |>{\centering\arraybackslash}p{2.5cm}|>{\centering\arraybackslash}p{2.5cm}|>{\centering\arraybackslash}p{2.5cm}| }
  \hline
  Homogeneous segments& Mean direction in radians (degrees) & Concentration \\

  \hline
 1-47	&	1.79(102.31)	&	0.9539627	\\
48-69	&	1.45(83.17)	&	0.9979347	\\
70-103	&	1.06(60.48)	&	0.9339317	\\
104-118	&	2.19(125.7)	&	0.9633238	\\
119-138	&	1.60(91.6)	&	0.9874869	\\
139-156	&	1.97(112.88)	&	0.978662	\\
157-168	&	2.42(138.77)	&	0.9729238	\\
169-235	&	4.12(236.06)	&	0.9787712	\\
236-258	&	3.74(214.08)	&	0.9941826	\\
  \hline
\end{tabular}
\vspace{0.3cm}
\caption{Values of the mean direction in radian (degree) and concentration for each segment where no change is detected by the SAGC test for the  Amphan cyclone data.}
\label{table:data_amphan_mean_concentration_table}
\end{table}



\newpage

\begin{figure}[h!]
    \centering
        {\includegraphics[width=0.6\textwidth, height=0.55\textwidth]{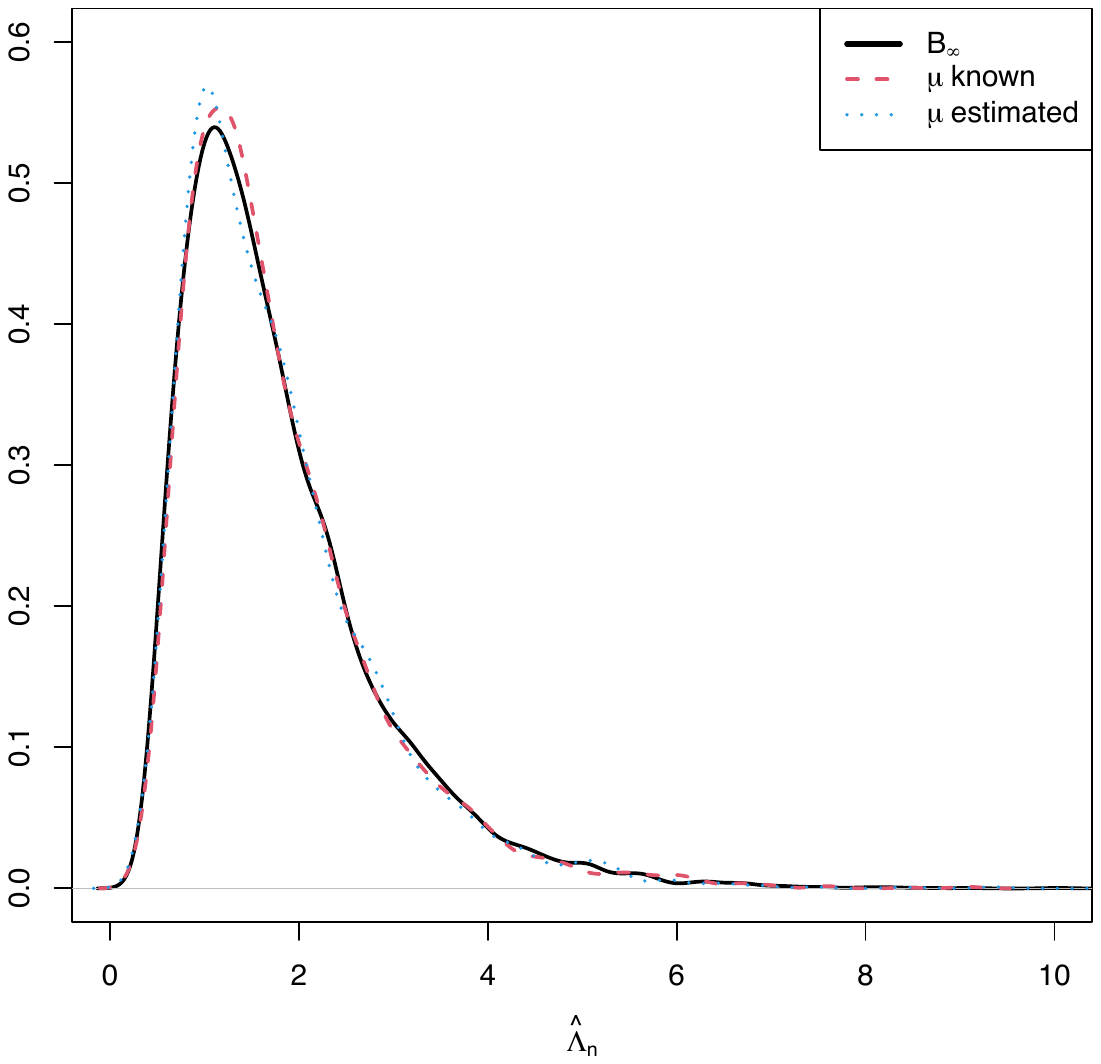}}\hspace{5pt}
  
    \caption{Density plot of the  test statistic, $\widehat{\Lambda}_n$ under $H_{0c}$  with sample of size $n=1000$ from von Mises distribution with the mean  $\mu$ is unknown 
 known and $\mu$ is estimated for the concentration parameter, $\kappa=1$ along with the density plot of the limiting random variable $B_{\infty} $.
 }
    \label{fig: density_sb_null_mean_unknown}
\end{figure}

\begin{figure}[h!]
    \centering
    \subfloat[ ]{%
        {\includegraphics[width=0.4\textwidth, height=0.35\textwidth]{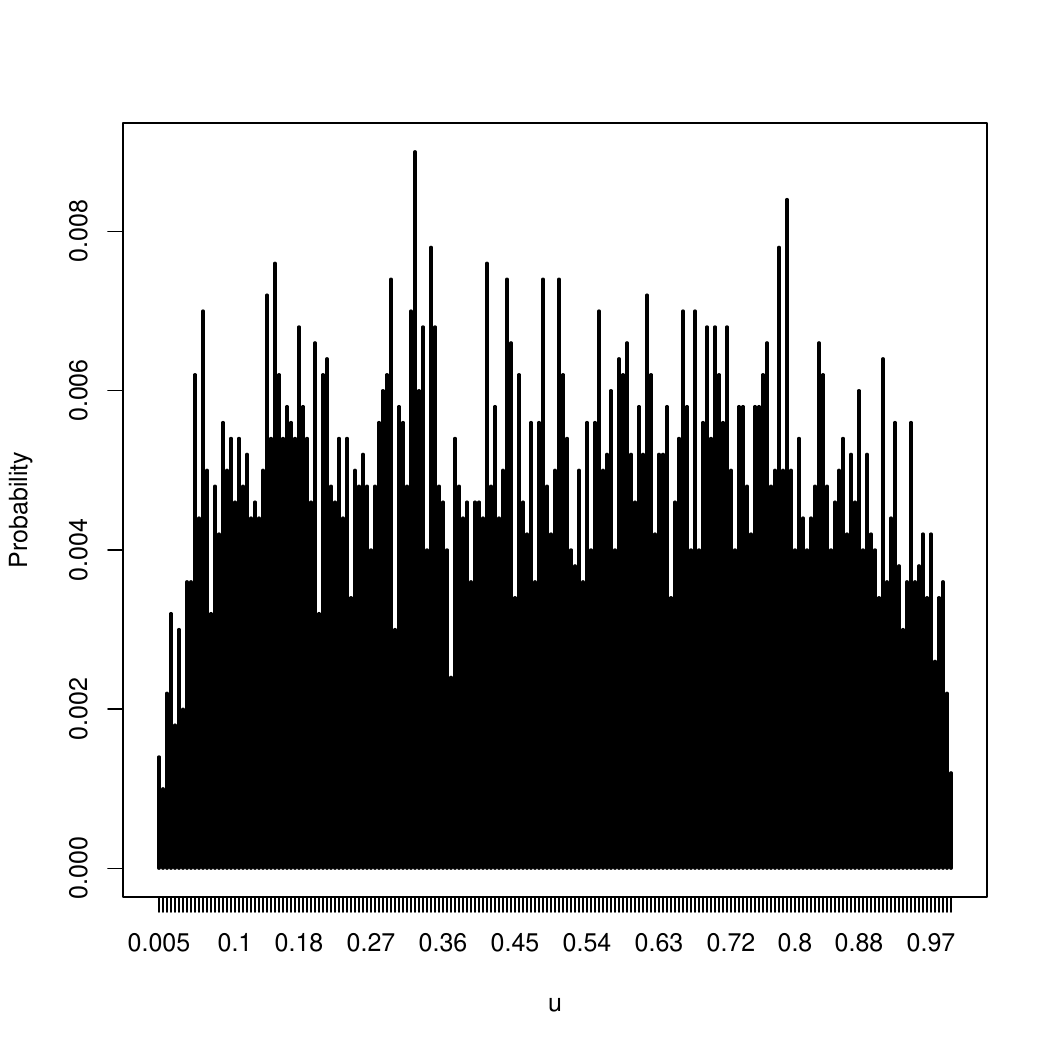}}}\hspace{5pt}
          \subfloat[]{%
        {\includegraphics[width=0.4\textwidth, height=0.35\textwidth]{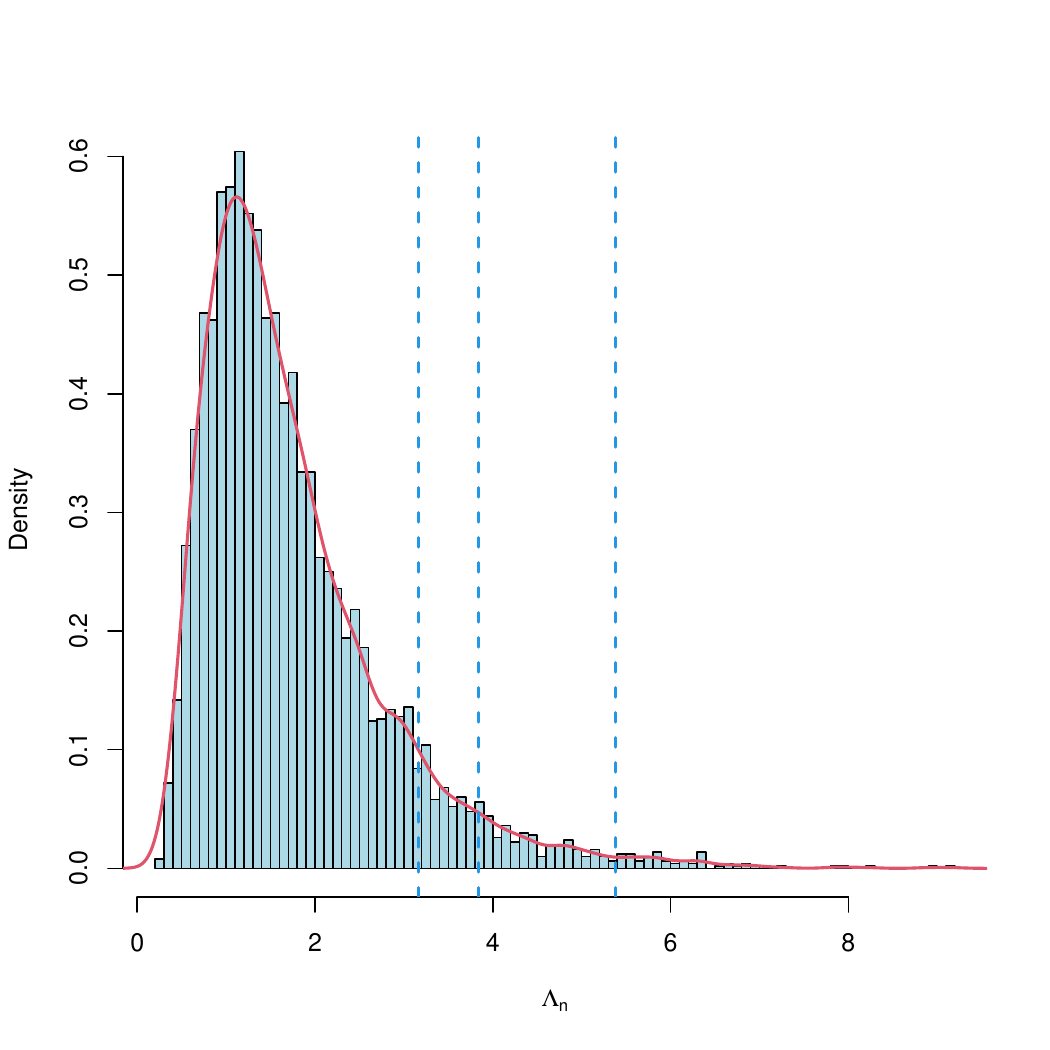}}}\hspace{5pt}
  
    \caption{(a) Distribution of the estimated location of changepoint in concentration parameter using SACC method. (b) Histogram of  SACC test statistic, $\Lambda_n$ along with the 0.90th, 0.95th, 0.99th quantiles under the null hypothesis, $H_{0c}$ when the data are from von Mises distribution (sample size $n=500$) with concentration parameter, $\kappa=1$ and the known mean direction $\mu=0$.}
    \label{sb_null_kappa_plot}
\end{figure}


\begin{figure}[h!]
    \centering
    \subfloat[ ]{%
        {\includegraphics[width=0.4\textwidth, height=0.35\textwidth]{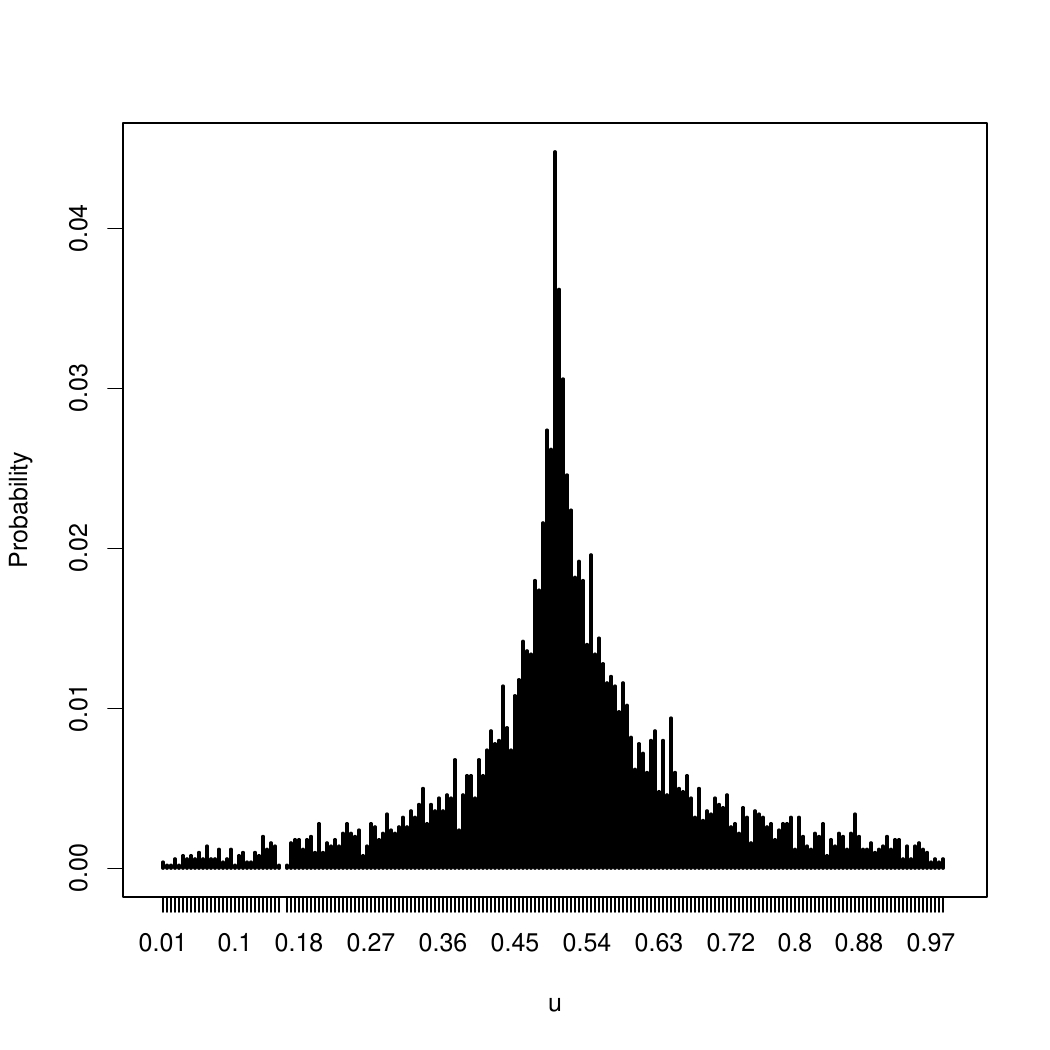}}}\hspace{5pt}
          \subfloat[ ]{%
        {\includegraphics[width=0.4\textwidth, height=0.35\textwidth]{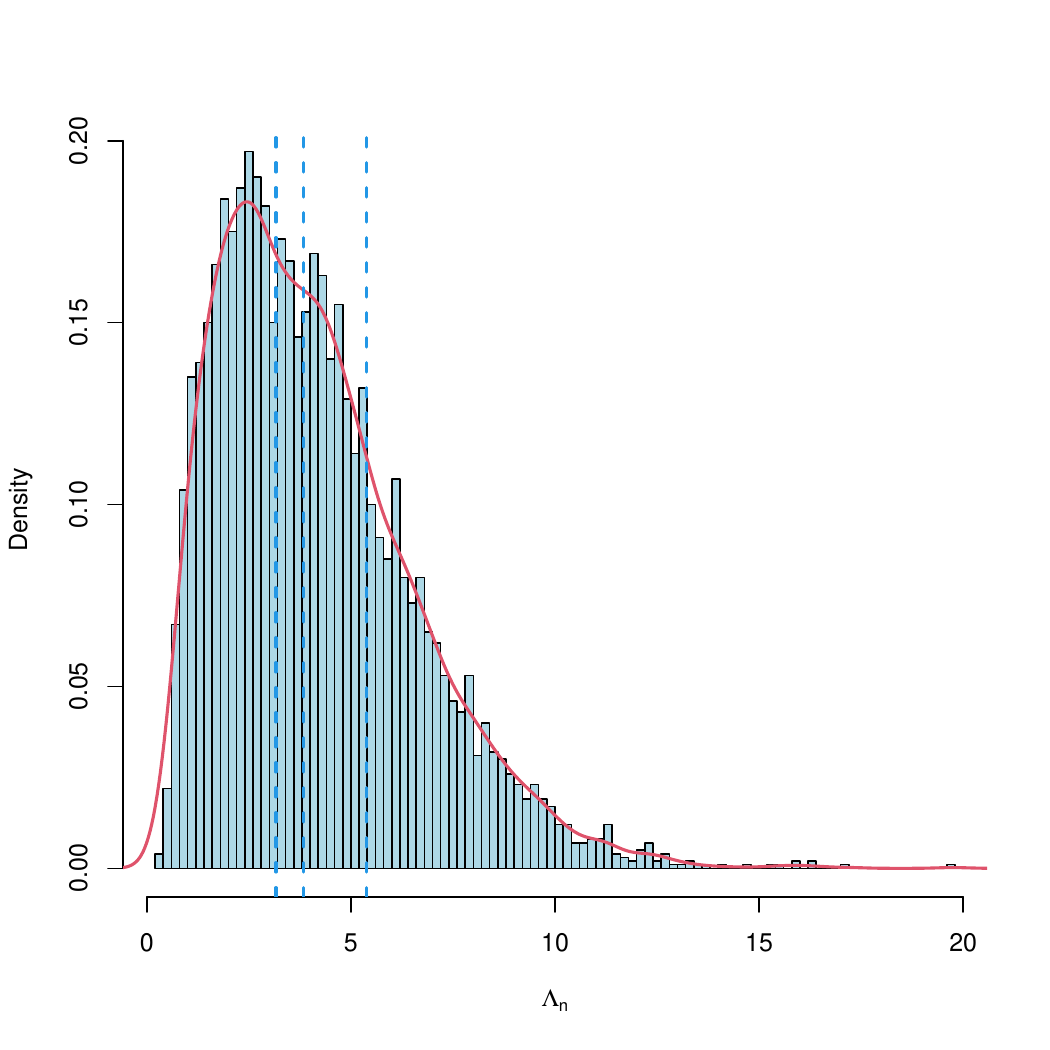}}}\hspace{5pt}
    
    \caption{(a) Distribution of the estimated location of changepoint in concentration parameter under the alternative hypothesis, $H_{1c}$ using SACC method. (b) Histogram of the SACC test statistic, $\Lambda_n$ when the true changepoint is at $k^*=\dfrac{n}{2}$,  $\kappa_1=1, \kappa_2=0.5 $ and mean direction $\mu=0$ under the alternative hypothesis, $H_{1c}$ along with the 0.90th, 0.95th, 0.99th quantiles when the data are from von Mises distribution (sample size $n=500$) under $H_{0c}$.}
    \label{sb_alt_kappa_plot}
\end{figure}


\newpage
\begin{figure}[h!]
    \centering
    \subfloat[ level $1\%$.]{%
        {\includegraphics[width=0.32\textwidth, height=0.32\textwidth]{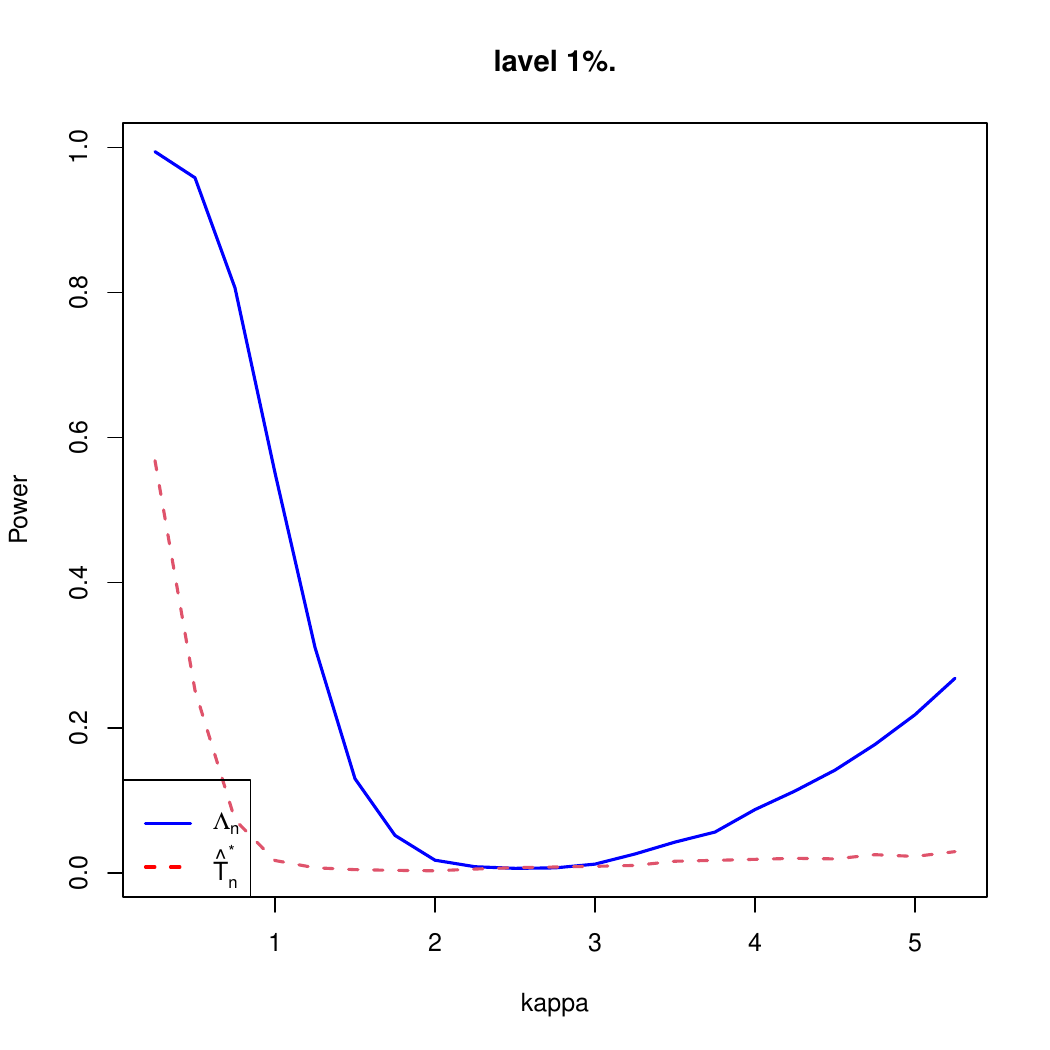}}}\hspace{5pt}
    \subfloat[ level $5\%$.]{%
        {\includegraphics[width=0.32\textwidth, height=0.32\textwidth]{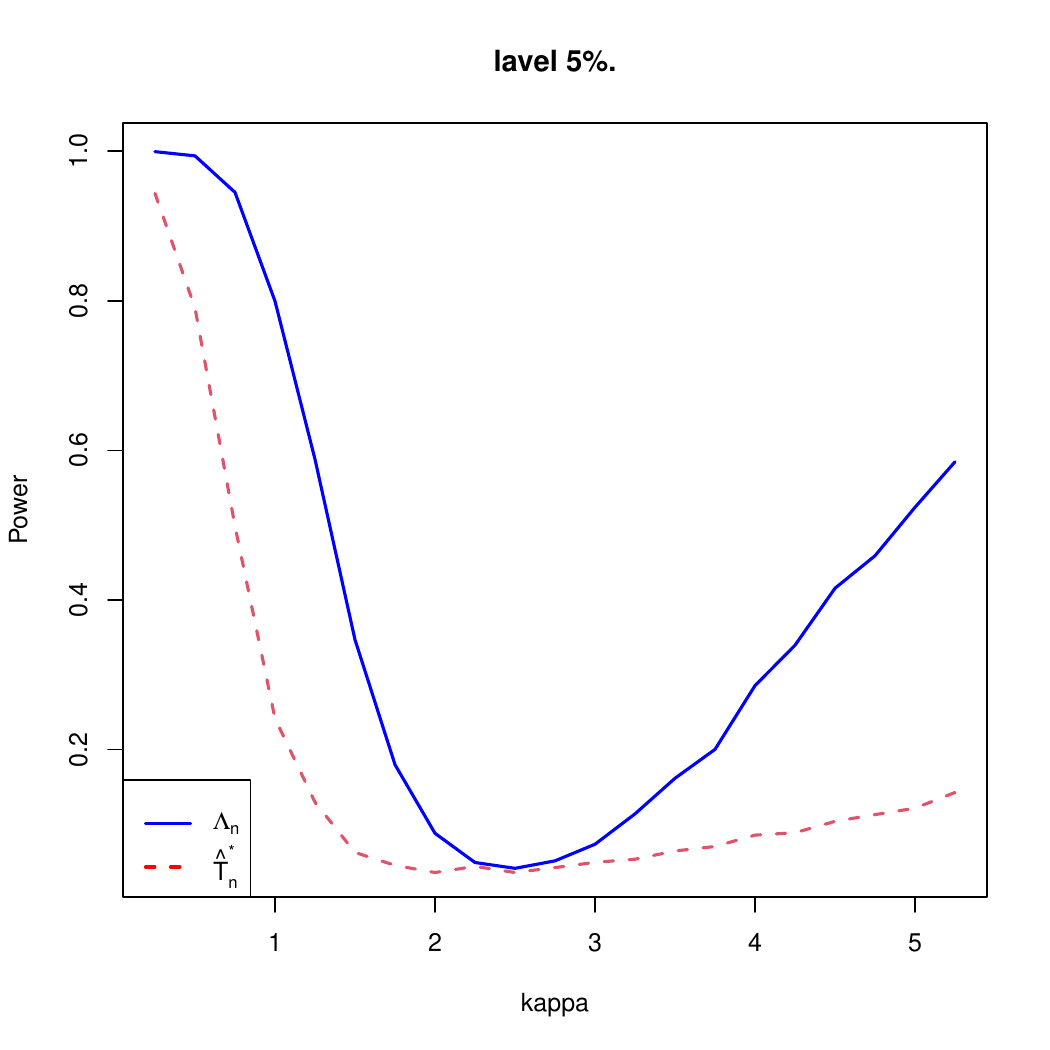}}}
    \vspace{5pt} 
     \caption{Power comparison between $\Lambda_n$ and $\widehat{T}_n^{*}$ with respect to $\kappa$ for the sample size of $n=100,$ drawn from von Mises distribution  with true changepoint  at $k^*=\dfrac{n}{2}$ under $H_{1c}.$ and $\mu=0, \kappa=2.5$ under $H_{0c}.$}
    \label{power_100ss}
\end{figure}


\begin{figure}[h!]
    \centering
    \subfloat[ level $1\%$  .]{%
        {\includegraphics[width=0.32\textwidth, height=0.32\textwidth]{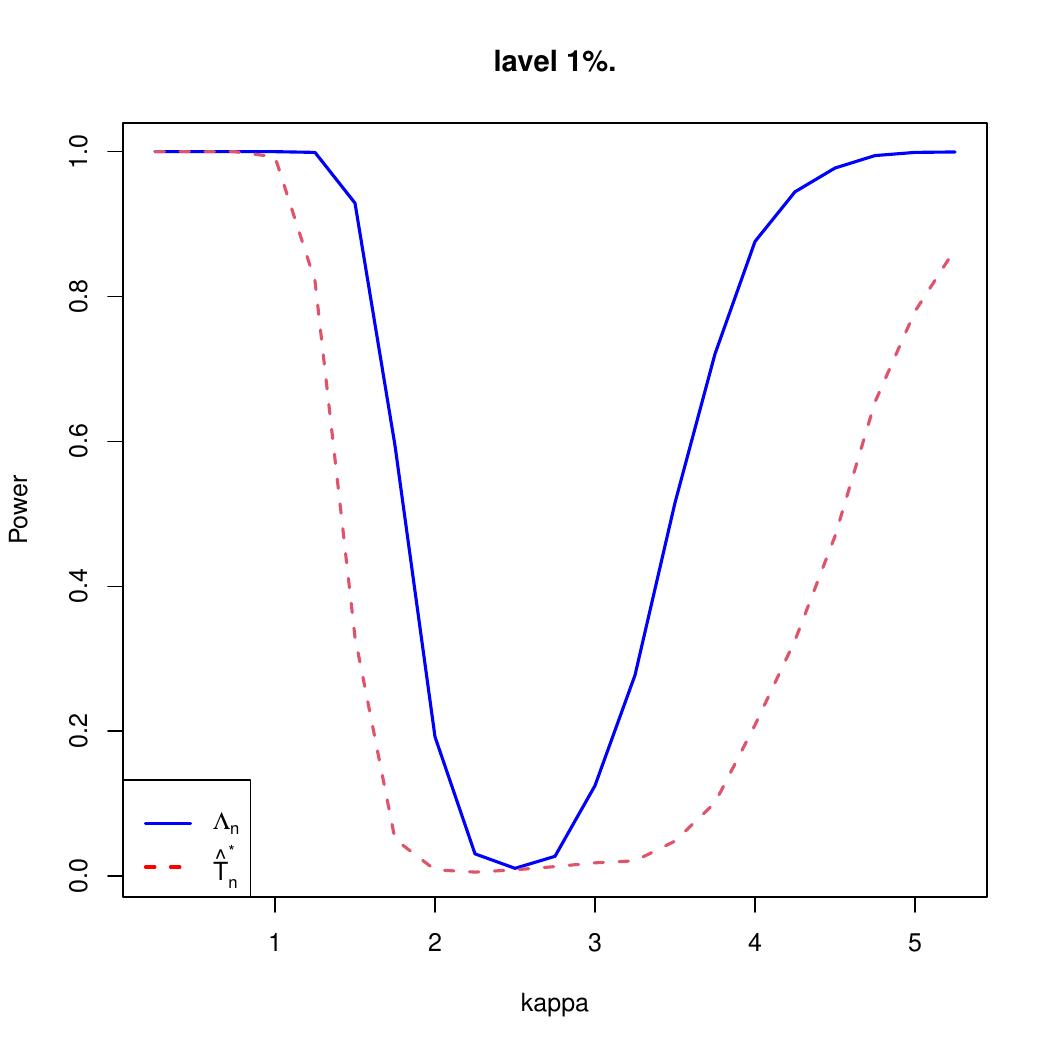}}}\hspace{5pt}
    \subfloat[ level $5\%$.]{%
        {\includegraphics[width=0.32\textwidth, height=0.32\textwidth]{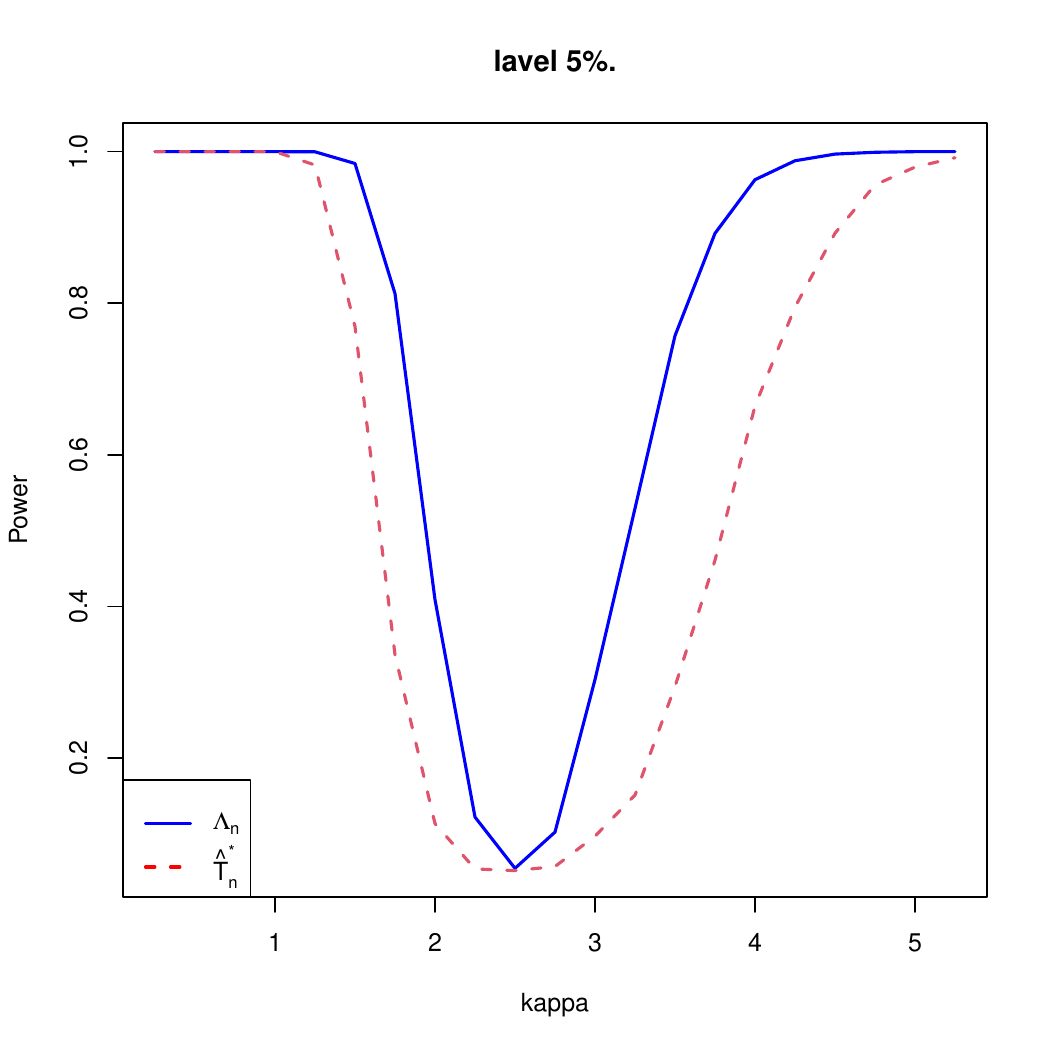}}}
    \vspace{5pt} 
    \caption{Power comparison between $\Lambda_n$ and $\widehat{T}_n^{*}$ with respect to $\kappa$ for the sample size of $n=500,$ drawn from von Mises distribution  with true changepoint  at $k^*=\dfrac{n}{2}$ under $H_{1c}.$ and $\mu=0, \kappa=2.5$ under $H_{0c}.$}
     \label{power_500ss}
\end{figure}

\begin{figure}[h!]
    \centering
    \subfloat[ level $1\%$  .]{%
        {\includegraphics[width=0.32\textwidth, height=0.32\textwidth]{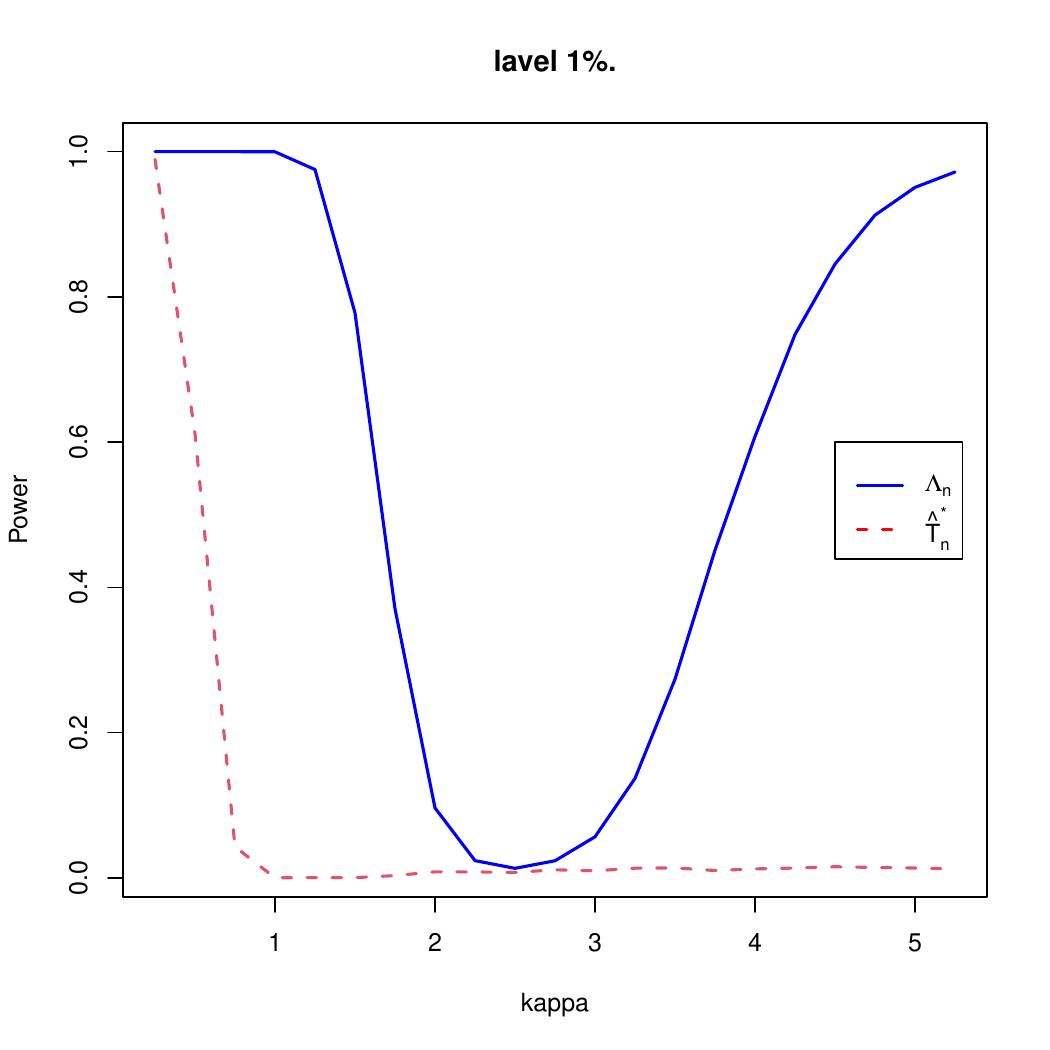}}}\hspace{5pt}
    \subfloat[ level $5\%$.]{%
        {\includegraphics[width=0.32\textwidth, height=0.32\textwidth]{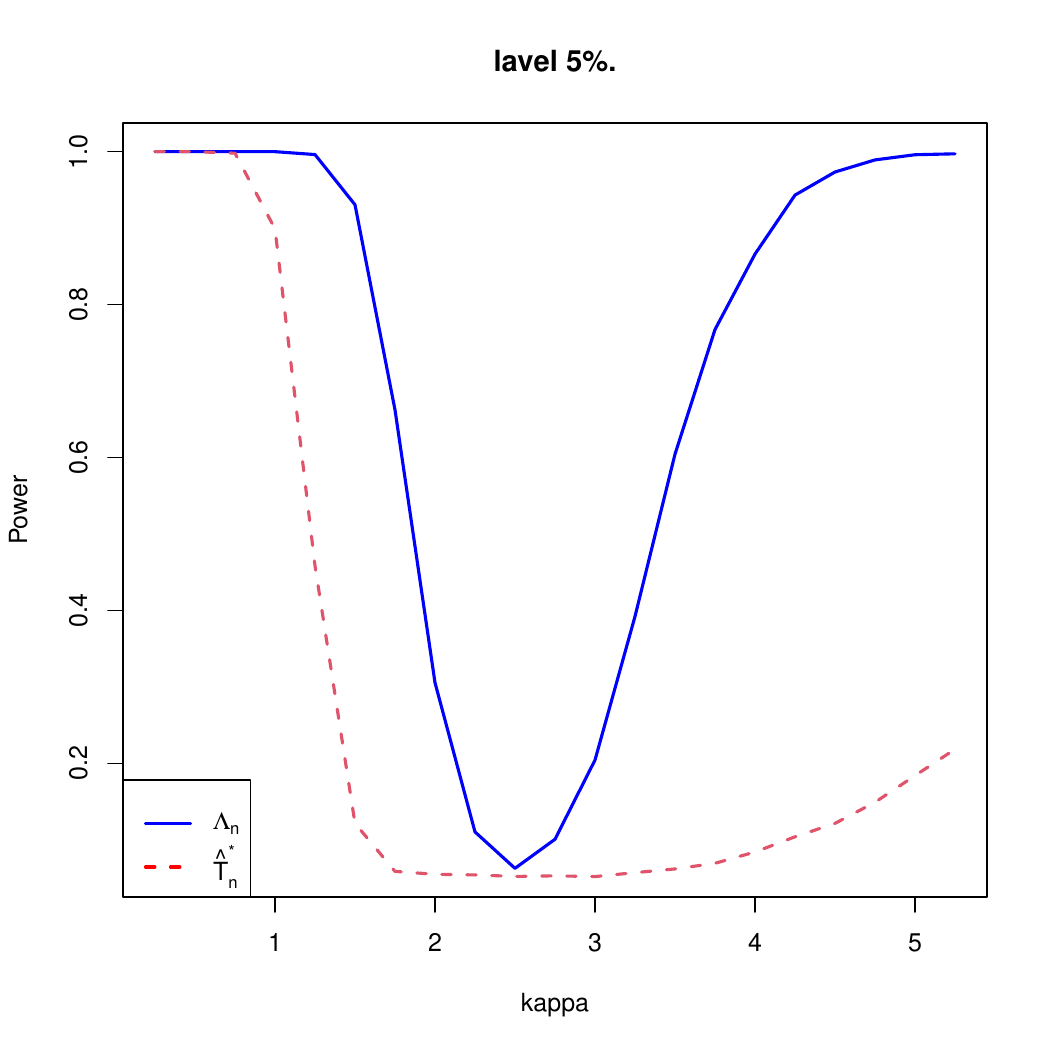}}}
    \vspace{5pt} 
    \caption{Power comparison between $\Lambda_n$ and $\widehat{T}_n^{*}$ with respect to $\kappa$ for the sample size of $n=500,$ when the data are from Kato-Jones distribution with true changepoint  at $k^*=\dfrac{n}{2}$ under $H_{1c}.$ and $\mu=\nu=0, \rho=0.4,\kappa=2.5 $ under $H_{0c}$.}
     \label{kjpower_kappa}
\end{figure}

\begin{figure}[h!]
    \centering
    \subfloat[ level $1\%$  .]{%
        {\includegraphics[width=0.32\textwidth, height=0.32\textwidth]{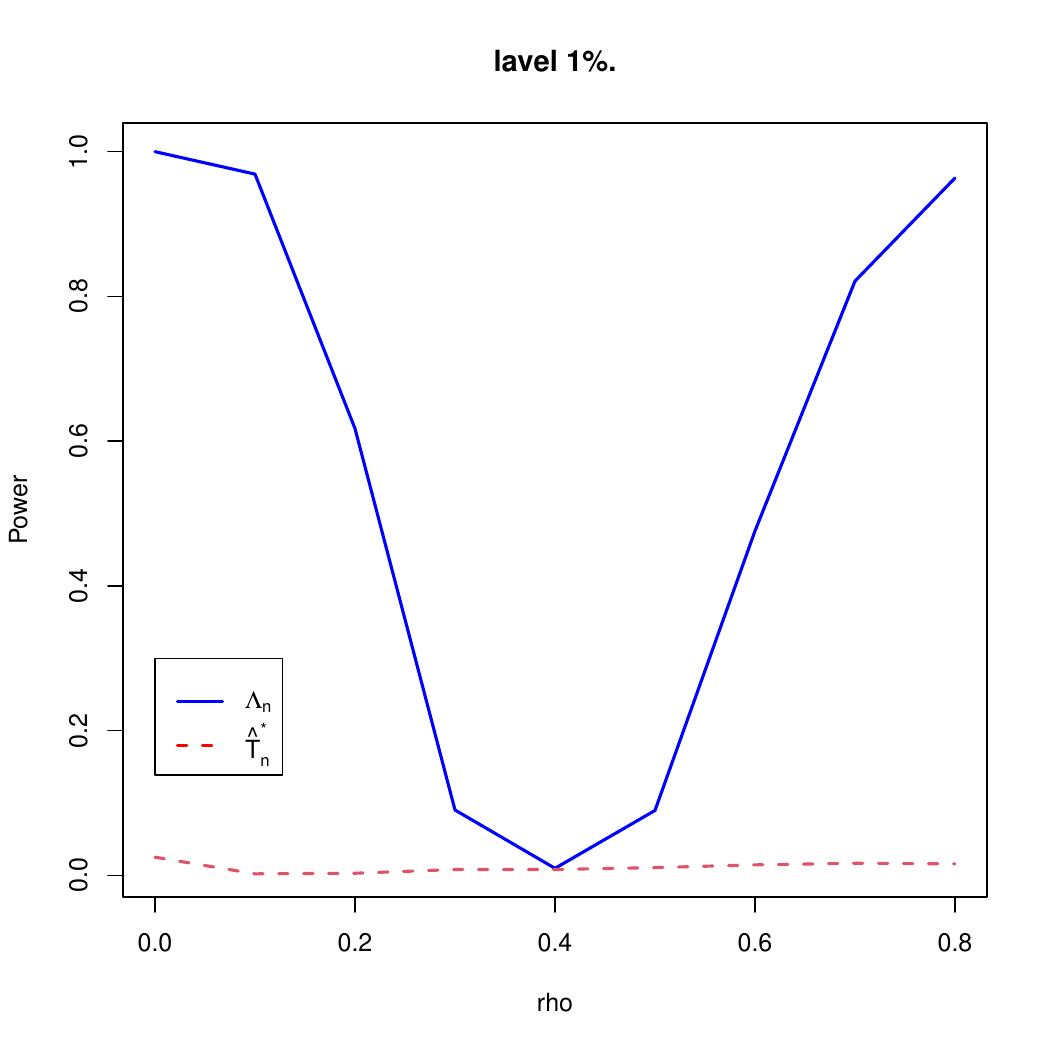}}}\hspace{5pt}
    \subfloat[ level $5\%$.]{%
        {\includegraphics[width=0.32\textwidth, height=0.32\textwidth]{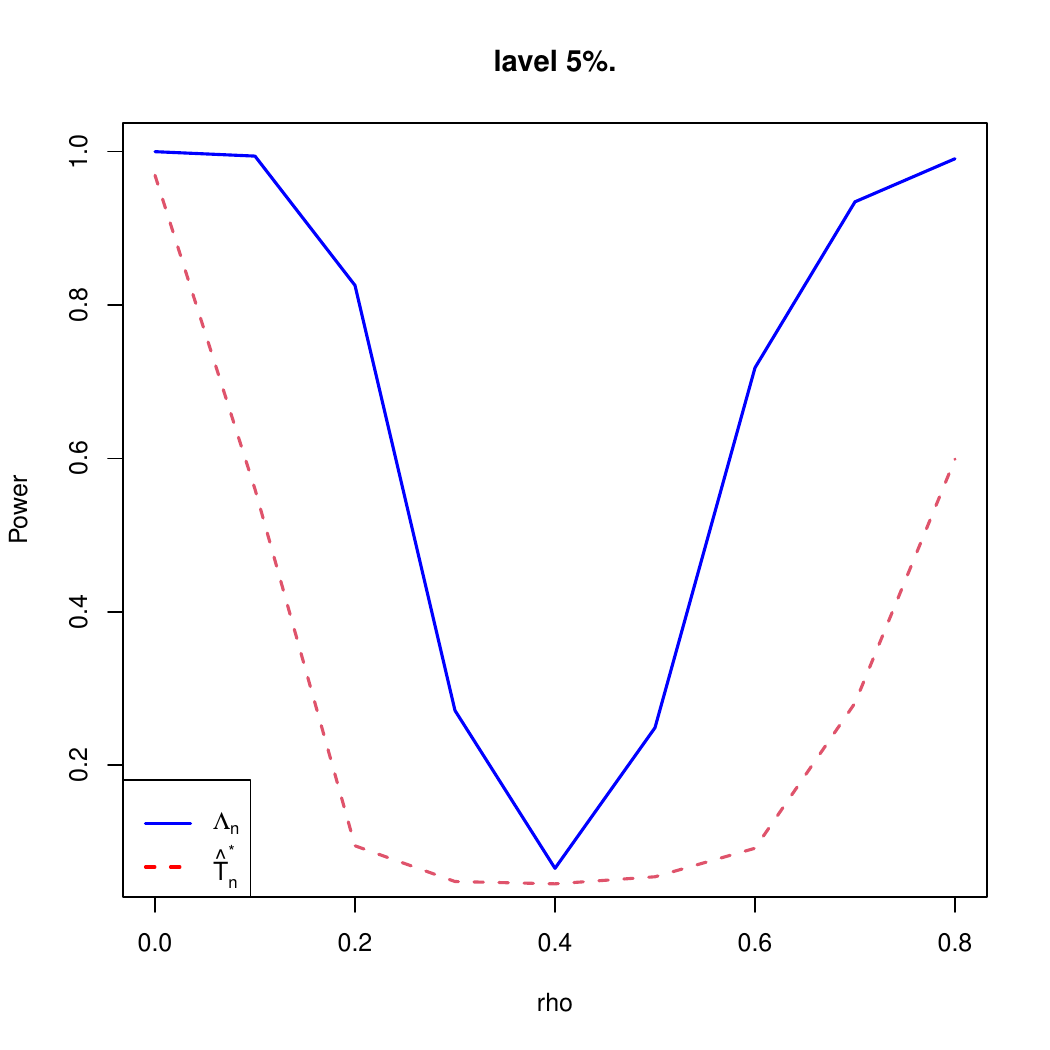}}}
    \vspace{5pt} 
    \caption{Power comparison between $\Lambda_n$ and $\widehat{T}_n^{*}$ with respect to $\rho$ for the sample size of $n=500,$ when the data are from Kato-Jones distribution with true changepoint at $k^*=\dfrac{n}{2}$ under $H_{1c}.$ and $\mu=\nu=0, \rho=0.4,\kappa=2.5 $ under $H_{0c}$.}
     \label{kjpower_r}
\end{figure}


\begin{figure*}[h!]
\centering
	\includegraphics[width=0.6\textwidth,height=0.4\textwidth]{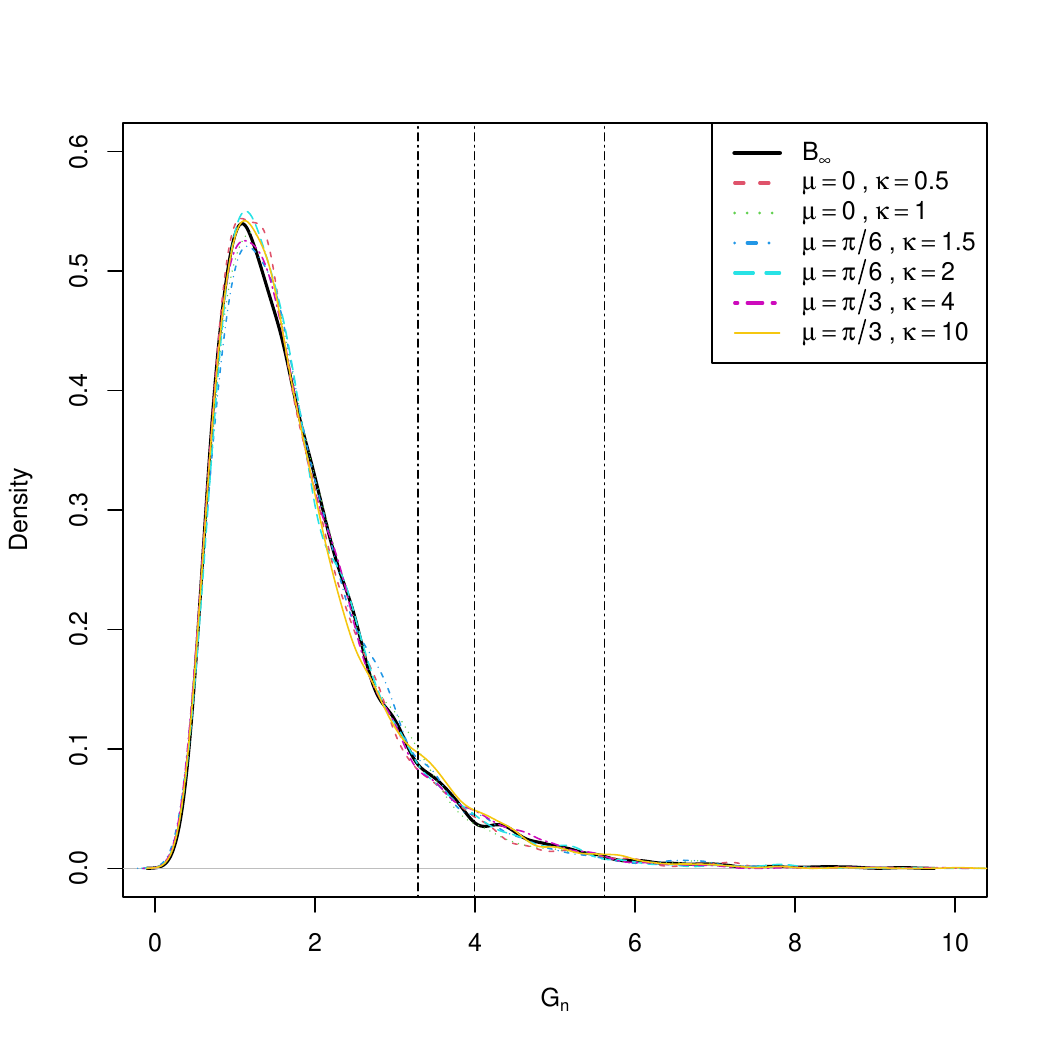}
	\caption{Density plot of the test statistic, $\mathcal{G}_n$ under $H_{0g}$  with a sample of size $n=1000$ from von Mises distribution with the different mean directions, and different concentration parameters along with density plot of the limiting random variable $B_{\infty} $, and its $0.90th,0.95th,$ and $0.99th$ quantiles.}
    \label{fig: density_general_null}
\end{figure*}

\begin{figure*}[h!]
\centering
	\includegraphics[width=0.6\textwidth,height=0.4\textwidth]{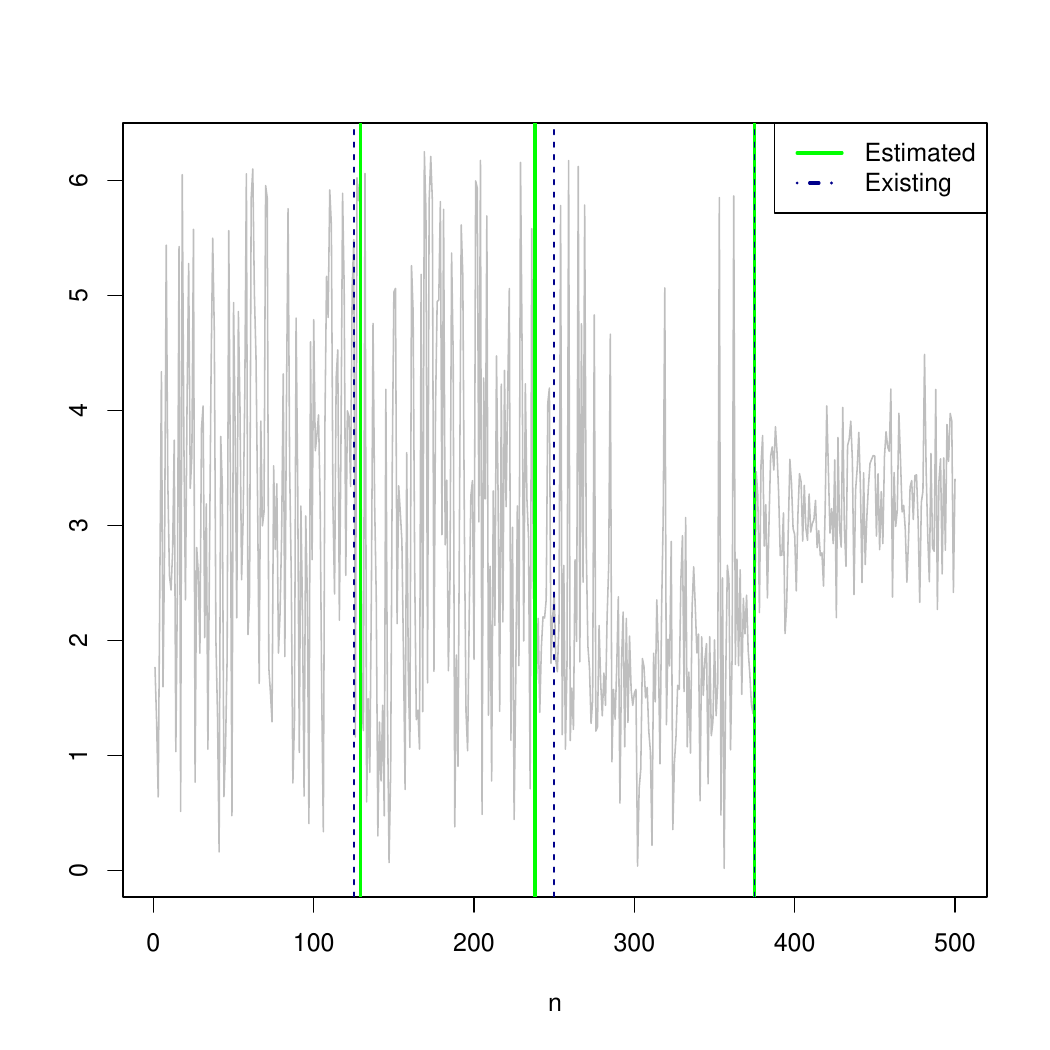}
	\caption{ Binary segmentation for changepoint detection in either or both the parameters of von Mises distribution. Existing changepoints are at $125~(\mu=\pi \mbox{~to~} \mu=\pi/2) $, $250~(\kappa=0.5 \mbox{~to~} \kappa=2) $, and $375~ (\mu=\pi/2,  \kappa=2 \mbox{~to~} \mu=\pi , \kappa=4) $. Estimated changepoints are at $129$, $238$, and $375$.}
    \label{fig: general_alt_plot}
\end{figure*}


\begin{figure*}[h]
	\centering
\includegraphics[width=0.6\textwidth,height=0.4\textwidth]{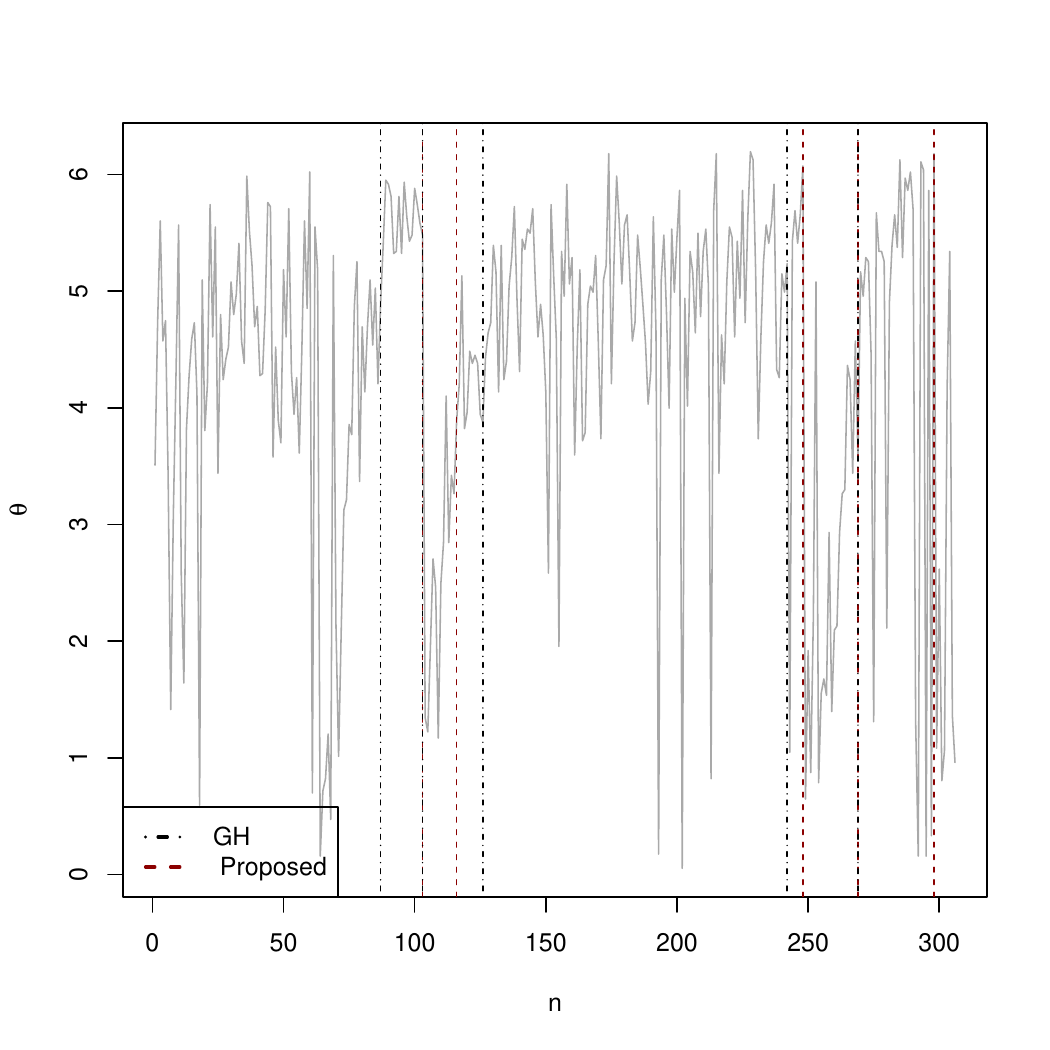}
	\caption{The plot of the Acrophase data with the estimated location of the changepoints in concentration, shown with the vertical lines, by the proposed method of SACC test and that of \cite{grabovsky2001change} denoted by GH.}
    \label{fig: data_analysis_plot}
\end{figure*}



\begin{figure*}[h]
\centering
	\includegraphics[width=0.6\textwidth,height=0.4\textwidth]{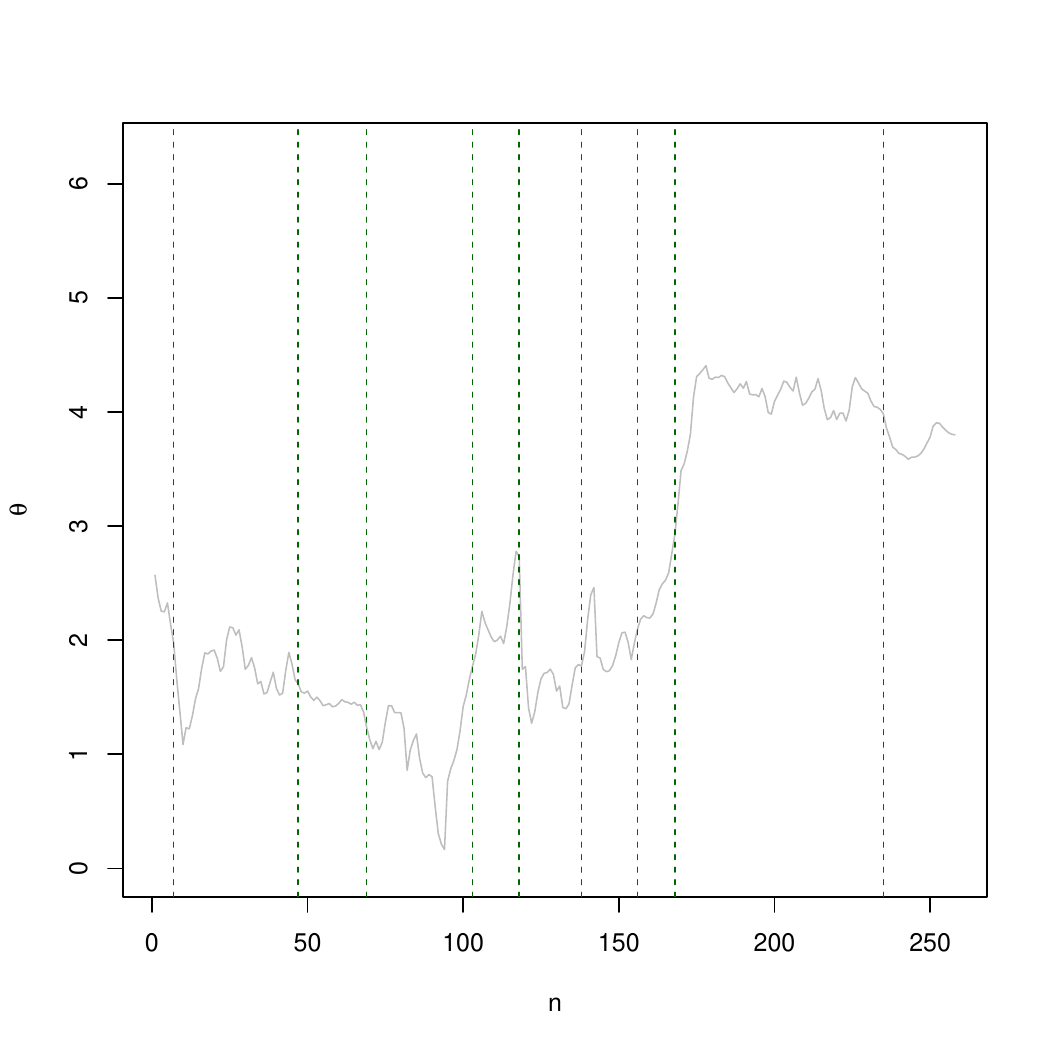}
	\caption{ The plot of the Super Cyclonic Storm (SuCS) ``AMPHAN'' data with the estimated location of the changepoints, shown with the vertical lines, by the proposed method of the SAGC test.}
    \label{fig: data_analysis_amphan_plot}
\end{figure*}


\label{algo}
\SetKwComment{Comment}{/* }{ */}
\RestyleAlgo{ruled}
\begin{algorithm}
\caption{Power calculation algorithm of the test statistic to detect changepoint in concentration parameters.}
\label{alg:algo_power_kappa}
$n \in \mathbb{N}$ \Comment*[r]{ Sample size}
$I \in \mathbb{N}$ \Comment*[r]{ Number of iteration, preferably large}
$R=r=1$  \Comment*[r]{ R,r are radius of unit circle}
$cp \in \{1,2,\ldots,(n-1) \}$ \Comment*[r]{Location of  true changepoint }
 $\kappa_0, \kappa_p \in (0, \infty)$ \Comment*[r]{In particular for von misses distribution}
 $\mu \in [ 0, 2\pi)$\;
  \For {$j =1,2, \ldots,I$}{

   $ \theta_1, \theta_2,\ldots, \theta_{cp} \sim f(\theta,\mu,\kappa_0) $ \;
   $ \theta_{cp+1}, \theta_{cp+2},\ldots, \theta_n \sim f(\theta,\mu,\kappa_p) $ \;

    $a_i\gets A_C^{(0)}[(\theta_i-\mu) \mod 2\pi]\mbox{~for~} i=1,2,\ldots,n$    \;
     $\widehat{Var(a)} \gets \frac{1}{n-1} \sum_{i=1}^{n}\left(a_i-\Bar{a}\right)^2, n\bar{a}=\displaystyle \sum_{i=1}^{n}a_i  $\;
    
       $T(k)\gets \frac{1}{n~\widehat{Var(a)}} \left[ \sum_{i=1}^{k} a_i-k\Bar{a}   \right]^2  \mbox{~~for all~~} k=1,\ldots,n $\;
      $\Lambda_n=\displaystyle \max_{1 \leq k < n}   \frac{T(k)}{\sqrt{\frac{k}{n}\left(1-\frac{k}{n} \right)}}$\;
    $Re[j] \gets \delta_{(\Lambda_n>l_{\alpha})}$ \Comment*[r]{$\delta$ is the Dirac delta function }
    
  }

$power \gets \displaystyle \frac{1}{I} \sum_{i=1}^{I}Re[i] $ \;
\KwResult{Power of the test can be obtained by varying the concentration parameter $k_p$ while keeping $k_0$ fixed.}
\end{algorithm}

\SetKwComment{Comment}{/* }{ */}
\RestyleAlgo{ruled}
\begin{algorithm}
\caption{Power calculation algorithm of the test statistic to detect changepoint in mean direction and/or concentration parameters.}
\label{alg:algo_general}
$n \in \mathbb{N}$ \Comment*[r]{ Sample size}
$I \in \mathbb{N}$ \Comment*[r]{ Number of iteration, preferably large}
$R=r=1$  \Comment*[r]{ R,r are radius of unit circle}
$cp \in \{1,2,\ldots,(n-1) \}$ \Comment*[r]{Location of  true changepoint }
 $\kappa_0, \kappa_p \in (0, \infty)$ \Comment*[r]{In particular for von misses distribution}
  $\mu_0, \mu_p \in (0, 2\pi]$ \;
  \For {$j =1,2, \ldots,I$}{

   $ \theta_1, \theta_2,\ldots, \theta_{cp} \sim f(\theta,\mu_0,\kappa_0) $ \;
   $ \theta_{cp+1}, \theta_{cp+2},\ldots, \theta_n \sim f(\theta,\mu_p,\kappa_p) $ \;

   $\hat{\mu}\in (0, 2\pi]$ \Comment*[r]{Estimated circular mean}

    $\Tilde{a}_i\gets \left[2\left( \delta_{(\theta_i<\pi)}-0.5 \right) \right] A_C^{(0)}(\theta_i) \mbox{~for~} i=1,2,\ldots,n$    \;
    
    $\hat{a}_i\gets A_C^{(0)}[(\theta_i-\hat{\mu}) \mod 2\pi]\mbox{~for~} i=1,2,\ldots,n$    \;
    
    $ \displaystyle d_i\gets\max_{1\leq i \leq n}\{\hat{a}_i,\Tilde{a}_i\}$\;
    
     $\widehat{Var(d)} \gets \frac{1}{n-1} \sum_{i=1}^{n}\left(d_i-\Bar{d}\right)^2, n\bar{d}=\displaystyle \sum_{i=1}^{n}d_i  $\;
    
       $U(k)\gets \frac{1}{n~\widehat{Var(d)}} \left[ \sum_{i=1}^{k} d_i-k\Bar{d}   \right]^2  \mbox{~~for all~~} k=1,\ldots,n $\;
      $\mathcal{G}_n=\displaystyle \max_{1 \leq k < n}   \frac{U(k)}{\sqrt{\frac{k}{n}\left(1-\frac{k}{n} \right)}}$\;
    $Re[j] \gets \delta_{(\mathcal{G}_n>g_{\alpha})}$ \Comment*[r]{$\delta$ is the Dirac delta function }
    
  }

$power \gets \displaystyle \frac{1}{I} \sum_{i=1}^{I}Re[i] $ \;
\KwResult{Power of the test can be obtained by varying the concentration parameter $k_p$ while keeping $k_0$ fixed.}
\end{algorithm}

\begin{table}[t]
\begin{center}
{\renewcommand{\arraystretch}{0.5}
  \begin{tabular}{cc|c|c|c|c|l}
\cline{3-5}
& & \multicolumn{3}{ c| }{Quantiles} \\ \cline{3-5}
& & 0.90th & 0.95th & 0.99th  \\ \cline{1-5}
\multicolumn{1}{ |c  }{\multirow{2}{*}{$n=50$} } &
\multicolumn{1}{ |c| }{$\kappa=0.5$} & 2.8664 & 3.5570& 5.1797   \\ \cline{2-5}
\multicolumn{1}{ |c  }{}                        &
\multicolumn{1}{ |c| }{$\kappa=1$}  & 2.8852 & 3.4706 & 4.9647    \\ \cline{2-5}
\multicolumn{1}{ |c  }{}                        &
\multicolumn{1}{ |c| }{$\kappa=1.5$} & 2.9324 & 3.4908  & 5.1092   \\ \cline{2-5}
\multicolumn{1}{ |c  }{}                        &
\multicolumn{1}{ |c| }{$\kappa=2$}  & 2.9059 & 3.5290 & 4.9475    \\ \cline{2-5}
\multicolumn{1}{ |c  }{}                        &
\multicolumn{1}{ |c| }{$\kappa=4$} & 2.9876  & 3.6723 & 5.2044   \\ \cline{2-5}
\multicolumn{1}{ |c  }{}                        &
\multicolumn{1}{ |c| }{$\kappa=10$} & 3.0115  & 3.6687 & 5.1525    \\ \cline{2-5}
\multicolumn{1}{ |c  }{}                        &
\multicolumn{1}{ |c| }{$B^{(50)}_\infty$} & 2.8967 & 3.5376 & 5.0784     \\ \cline{1-5}

\multicolumn{1}{ |c  }{\multirow{2}{*}{$n=100$} } &
\multicolumn{1}{ |c| }{$\kappa=0.5$} & 2.9655&	3.6087&	5.0154   \\ \cline{2-5}
\multicolumn{1}{ |c  }{}                        &
\multicolumn{1}{ |c| }{$\kappa=1$}  & 2.9998&	3.6626&	5.1375    \\ \cline{2-5}
\multicolumn{1}{ |c  }{}                        &
\multicolumn{1}{ |c| }{$\kappa=1.5$} & 2.9601&	3.5823&	5.0707    \\ \cline{2-5}
\multicolumn{1}{ |c  }{}                        &
\multicolumn{1}{ |c| }{$\kappa=2$}  & 3.0054 & 3.6686 & 5.1957   \\ \cline{2-5}
\multicolumn{1}{ |c  }{}                        &
\multicolumn{1}{ |c| }{$\kappa=4$} & 3.1320 & 3.8628 & 5.4352   \\ \cline{2-5}
\multicolumn{1}{ |c  }{}                        &
\multicolumn{1}{ |c| }{$\kappa=10$} & 3.0528 & 3.7376 & 5.1318    \\ \cline{2-5}
\multicolumn{1}{ |c  }{}                        &
\multicolumn{1}{ |c| }{$B^{(100)}_\infty$} & 2.9987&	3.6939&	5.2307     \\ \cline{1-5}

\multicolumn{1}{ |c  }{\multirow{2}{*}{$n=200$} } &
\multicolumn{1}{ |c| }{$\kappa=0.5$} & 3.0294&  3.6653&	5.1041     \\ \cline{2-5}
\multicolumn{1}{ |c  }{}                        &
\multicolumn{1}{ |c| }{$\kappa=1$}  & 3.0619&	3.7477&	5.3283    \\ \cline{2-5}
\multicolumn{1}{ |c  }{}                        &
\multicolumn{1}{ |c| }{$\kappa=1.5$} &3.0533&	3.7104&	5.0903  \\ \cline{2-5}
\multicolumn{1}{ |c  }{}                        &
\multicolumn{1}{ |c| }{$\kappa=2$}  & 3.1221 & 3.8160 &5.5242   \\ \cline{2-5}
\multicolumn{1}{ |c  }{}                        &
\multicolumn{1}{ |c| }{$\kappa=4$} & 3.1010& 3.8392& 5.1764   \\ \cline{2-5}
\multicolumn{1}{ |c  }{}                        &
\multicolumn{1}{ |c| }{$\kappa=10$} & 3.1391 & 3.8052& 5.3281   \\ \cline{2-5}
\multicolumn{1}{ |c  }{}                        &
\multicolumn{1}{ |c| }{$B^{(200)}_\infty$} & 3.0353& 3.6733&	5.2212   \\ \cline{1-5}

\multicolumn{1}{ |c  }{\multirow{2}{*}{$n=500$} } &
\multicolumn{1}{ |c| }{$\kappa=0.5$} & 3.1820 &	3.8620& 	5.3930     \\ \cline{2-5}
\multicolumn{1}{ |c  }{}                        &
\multicolumn{1}{ |c| }{$\kappa=1$}  & 3.1469&	3.9033& 	5.7212      \\ \cline{2-5}
\multicolumn{1}{ |c  }{}                        &
\multicolumn{1}{ |c| }{$\kappa=1.5$} & 3.2189&  3.8676& 	5.4404  \\ \cline{2-5}
\multicolumn{1}{ |c  }{}                        &
\multicolumn{1}{ |c| }{$\kappa=2$}  & 3.1148& 3.8149& 5.5284   \\ \cline{2-5}
\multicolumn{1}{ |c  }{}                        &
\multicolumn{1}{ |c| }{$\kappa=4$} & 3.0986 & 3.8050& 5.3250   \\ \cline{2-5}
\multicolumn{1}{ |c  }{}                        &
\multicolumn{1}{ |c| }{$\kappa=10$} & 3.1611 & 3.8389 & 5.4520   \\ \cline{2-5}
\multicolumn{1}{ |c  }{}                        &
\multicolumn{1}{ |c| }{$B^{(500)}_\infty$} &3.2224 &	3.9021 &  5.7649   \\ \cline{1-5}

\multicolumn{1}{ |c  }{\multirow{2}{*}{$n=1000$} } &
\multicolumn{1}{ |c| }{$\kappa=0.5$} &3.2093&	3.9429&	5.5770     \\ \cline{2-5}
\multicolumn{1}{ |c  }{}                        &
\multicolumn{1}{ |c| }{$\kappa=1$}  &3.2860&	3.9551& 	5.5877    \\ \cline{2-5}
\multicolumn{1}{ |c  }{}                        &
\multicolumn{1}{ |c| }{$\kappa=1.5$} & 3.2012 & 3.9068&  5.7190  \\ \cline{2-5}
\multicolumn{1}{ |c  }{}                        &
\multicolumn{1}{ |c| }{$\kappa=2$}  & 3.2164 & 3.9404 &5.4612   \\ \cline{2-5}
\multicolumn{1}{ |c  }{}                        &
\multicolumn{1}{ |c| }{$\kappa=4$} & 3.1152 &3.8271 &5.4818  \\ \cline{2-5}
\multicolumn{1}{ |c  }{}                        &
\multicolumn{1}{ |c| }{$\kappa=10$} & 3.2164 & 3.9404 & 5.4612    \\ \cline{2-5}
\multicolumn{1}{ |c  }{}                        &
\multicolumn{1}{ |c| }{$B^{(1000)}_\infty$} & 3.2173& 3.8994&	5.3781    \\ \cline{1-5}

\end{tabular}}
\end{center} \vspace{0.3cm}
\caption{Table of the cut-off values of the SACC test statistic $\Lambda_n$ under the null hypothesis, $H_{0c}$ when the sample of sizes of $50, 100,200, 500,$ and $1000$ are drawn from von Mises distribution with mean direction $\mu=0,$ and different concentration parameters $\kappa=0.5,1, 1.5,2,4$ and $10$. The table also contains the cut-off values from the limiting distribution of $B_\infty$ (Equation-\ref{bbridge}) with the grid size of $50,100,200, 500,$ and $1000$, respectively.}
\label{table:cut-off for concentration}
\end{table}

\end{document}